\documentclass{article}

\usepackage{arxiv}

\usepackage[utf8]{inputenc} 
\usepackage[T1]{fontenc}    
\usepackage{hyperref}       
\usepackage{url}            
\usepackage{booktabs}       
\usepackage{amsfonts}       
\usepackage{nicefrac}       
\usepackage{microtype}      
\usepackage{lipsum}
\usepackage{graphicx}
\graphicspath{ {./images/} }

\usepackage{framed,multirow}

\usepackage{amssymb}
\usepackage{latexsym}

\usepackage{tabu}                      
\usepackage{booktabs}                  
\usepackage{lipsum}                    
\usepackage{mwe}                       

\usepackage{comment}
\usepackage[table,svgnames]{xcolor}
\usepackage{subcaption}
\usepackage{amssymb}
\usepackage[utf8]{inputenc}
\usepackage[T1]{fontenc}
\usepackage{amsmath,amsfonts}
\usepackage{booktabs}
\usepackage{multirow}
\usepackage{tabularx}

\newcommand{\etal}{\textit{et al.~}}
\usepackage{mathptmx}   

\usepackage{url}
\usepackage{xcolor}
\definecolor{newcolor}{rgb}{.8,.349,.1}
\usepackage{ulem}

\usepackage{hyperref}
\usepackage{xr}
\usepackage{cleveref}
\usepackage[switch,pagewise]{lineno} 

\usepackage{tabu}                      
\usepackage{booktabs}                  

\usepackage{comment}
\usepackage[table,svgnames]{xcolor}
\usepackage{subcaption}
\usepackage{amssymb}
\usepackage[utf8]{inputenc}
\usepackage[T1]{fontenc}
\usepackage{amsmath,amsfonts}
\usepackage{booktabs,hhline}
\usepackage{multirow}
\usepackage{tabularx}

\usepackage{mathptmx}   

\usepackage{url}
\usepackage{xcolor}
\definecolor{newcolor}{rgb}{.8,.349,.1}

\usepackage{hyperref}

\usepackage[switch,pagewise]{lineno} 

\title{Time-Variant Vector Field Visualization for Magnetic Fields of Neutron Star Simulations}%

\author{
 Simon J. Lieb, William Cook, Jan Hombeck, Sebastiano Bernuzzi, Kai Lawonn \\
  Friedrich-Schiller-University Jena\\ 
  Jena, Germany\\
}

\begin{document}
\maketitle
\begin{abstract}
We present a novel visualization application designed to explore the time-dependent development of magnetic fields of neutron stars.
The strongest magnetic fields in the universe can be found within neutron stars, potentially playing a role in initiating astrophysical jets and facilitating the outflow of neutron-rich matter, ultimately resulting in the production of heavy elements during binary neutron star mergers.
Since such effects may be dependent on the strength and configuration of the magnetic field, the formation and parameters of such fields are part of current research in astrophysics.
Magnetic fields are investigated using simulations in which various initial configurations are tested. 
However, the long-term configuration is an open question, and current simulations do not achieve a stable magnetic field.
Neutron star simulations produce data quantities in the range of several terabytes, which are both spatially in 3D and temporally resolved.
Our tool enables physicists to interactively explore the generated data.
We first convert the data in a pre-processing step and then we combine sparse vector field visualization using streamlines with dense vector field visualization using line integral convolution.
We provide several methods to interact with the data responsively.
This allows the user to intuitively investigate data-specific issues.
Furthermore, diverse visualization techniques facilitate individual exploration of the data and enable real-time processing of specific domain tasks, like the investigation of the time-dependent evolution of the magnetic field.
In a qualitative study, domain experts tested the tool, and the usability was queried.
Experts rated the tool very positively and recommended it for their daily work.
\end{abstract}


\section{Introduction}

At the latest after the detection of the gravitational wave signal GW170817 on 17 August 2017 \cite{LIGOScientific:2017vwq}, the interest and importance of research into neutron star properties has grown steadily, as this event had far-reaching consequences for the understanding of astrophysics and the formation of matter in the universe.
At the same time, the importance of visualizing the correspondingly complex data is growing. In this paper, we present our visualization tool, which is specially adapted to the complex physical data described below.
\textit{Neutron Stars} (NSs) can contain some of the strongest magnetic fields occurring in the universe, up to $10^{16}$Gauss.
A system of two co-orbiting NSs, called \textit{Binary neutron star systems} (BNS), lose energy over time due to gravitational wave emission, eventually merging and forming a remnant NS or collapsing into a remnant black hole.
Such a merger gives rise to a wide range of observable astrophysical phenomena, for instance an observed burst of gamma rays \cite{Goldstein:2017mmi,Savchenko:2017ffs} believed to originate from the post-merger remnant launching out high velocity ($ > 99\% $ of the speed of light) matter, known as a jet \cite{Paczynski:1986px}. 
It has been suggested that matter ejected during such a BNS merger will undergo radioactive decays that produce a large proportion of the heavy chemical elements found in the universe, which long term observations of the above BNS merger support \cite{Kasen:2017a}.

\begin{figure}[t]

    \begin{subfigure}{.33\linewidth}
  \centering
    \includegraphics[width=1\linewidth]{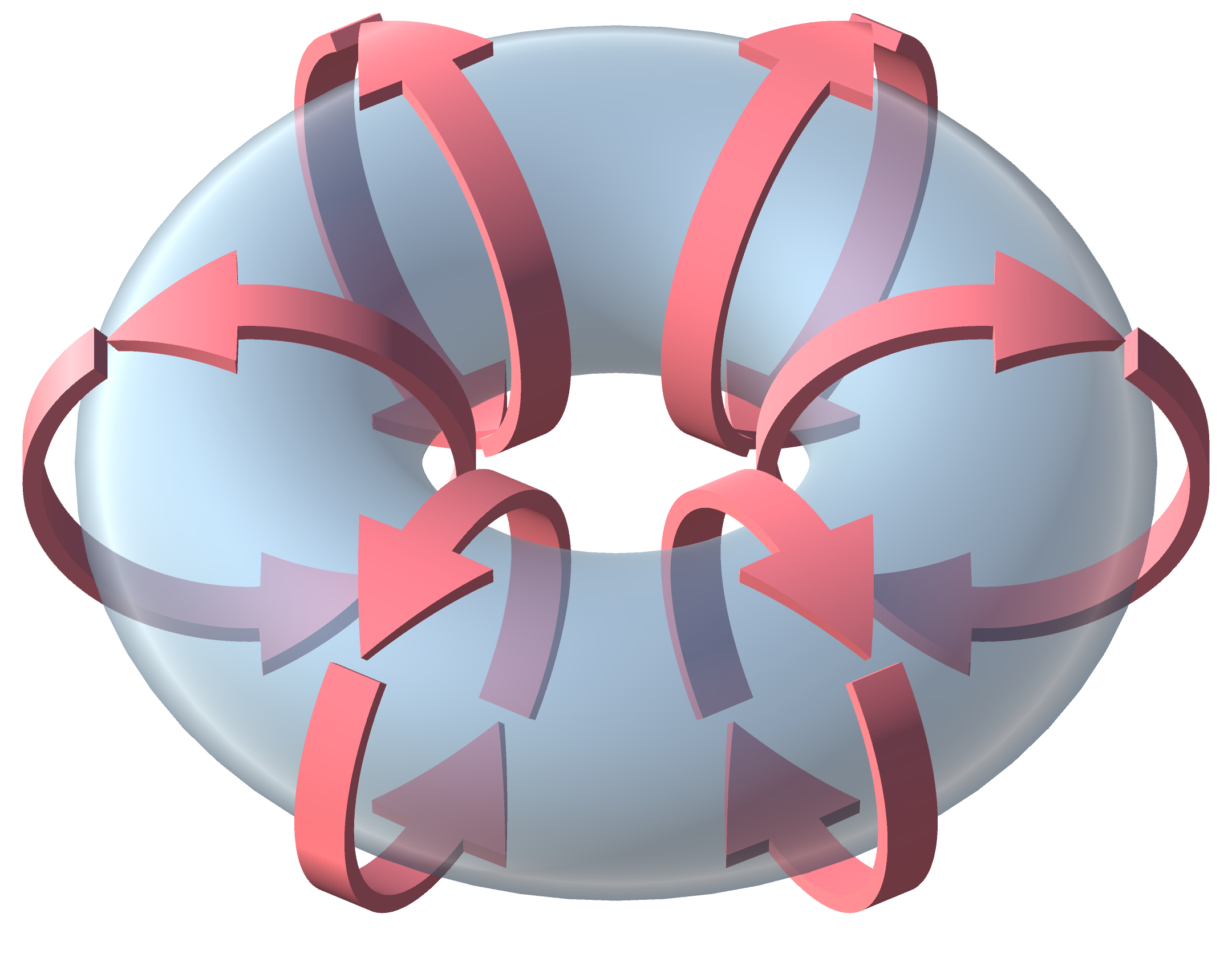}
    \caption{poloidal}
  \end{subfigure}%
  \begin{subfigure}{.33\linewidth}
  \centering
    \includegraphics[width=1\linewidth]{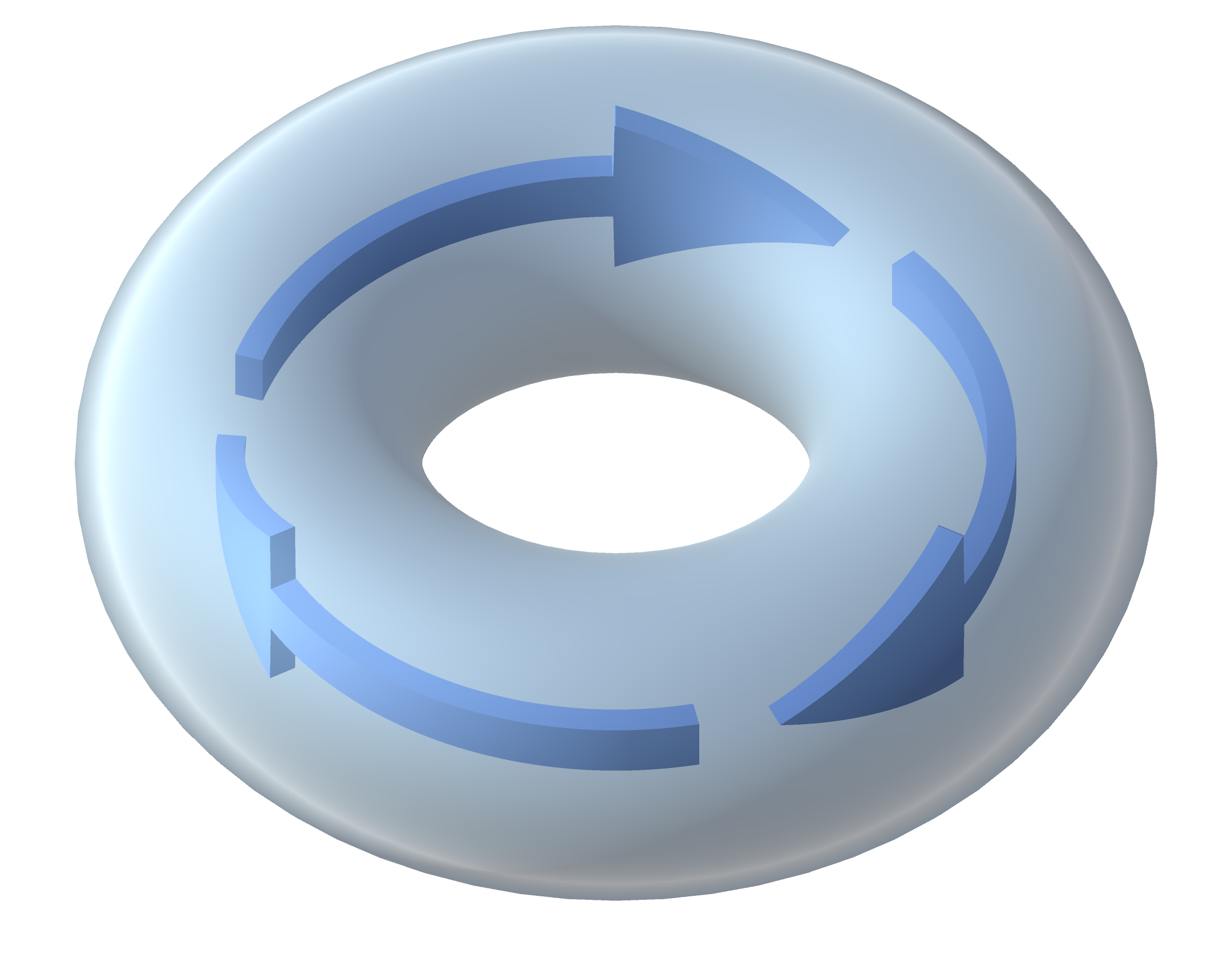}
    \caption{toroidal}
  \end{subfigure}
  \begin{subfigure}{.33\linewidth}
  \centering
    \includegraphics[width=1\linewidth]{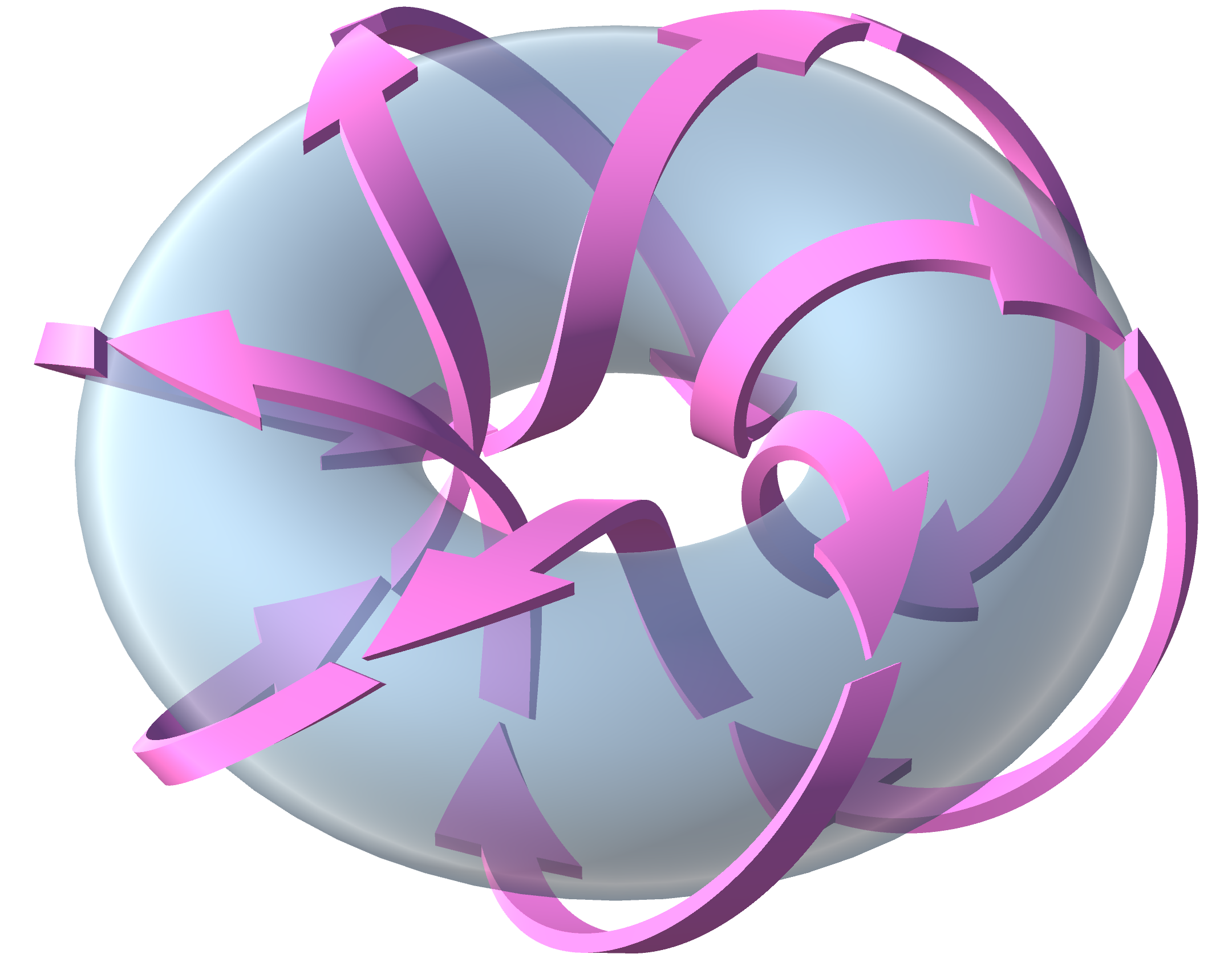}
    \caption{poloidal-toroidal mixed}
  \end{subfigure}
  
 \caption{Configuration of magnetic fields in neutron stars. (a) poloidal, (b) toroidal, (c) poloidal and toroidal mixed}
 \label{Fig:FieldConfig}
\end{figure}

Understanding the formation of jets and matter ejection in BNS mergers is influenced by magnetic fields during and after the merger \cite{Rezzolla:2011da,Ruiz:2016rai,Siegel:2014a,Kiuchi:2014hja}. 
The configuration of magnetic fields within neutron stars is crucial for modeling these astrophysical processes. 
However, the exact magnetic field configuration in neutron stars remains an open question, with certain configurations experiencing instabilities \cite{Tayler:1957a,Tayler:1973a,Markey:1973a,Markey:1974a}. 
Studying the long-term evolution of these fields can provide insights into configurations expected in neutron stars undergoing BNS mergers. 
Findings about configurations are gained by visualizations and exploring the data obtained. The current process of visualization is cumbersome, which is why a special tool for this data proves to be very useful.
Usually, magnetic fields of neutron stars consist of a mixture of poloidal and toroidal configurations, see Fig. \ref{Fig:FieldConfig}.
A recent study explored configurations with initially poloidal fields; that is, fields which emanate axisymmetrically from the north pole of the neutron star and point towards the south pole, similar to the magnetic field of a simple bar magnet. This revealed the development of toroidal components of the magnetic field; that is components perpendicular to the poloidal field, which wind around the equator of the neutron star, but with persistent energy loss in the magnetic field sector at the simulation's end \cite{sur2022long}. 
The simulations generate extensive data, including volumetric representations of mass, vector fields, and structural data for each time step, posing challenges for analysis with generic tools due to its size.

\begin{figure*}[t]
    \centering
\includegraphics[width=0.7\linewidth]{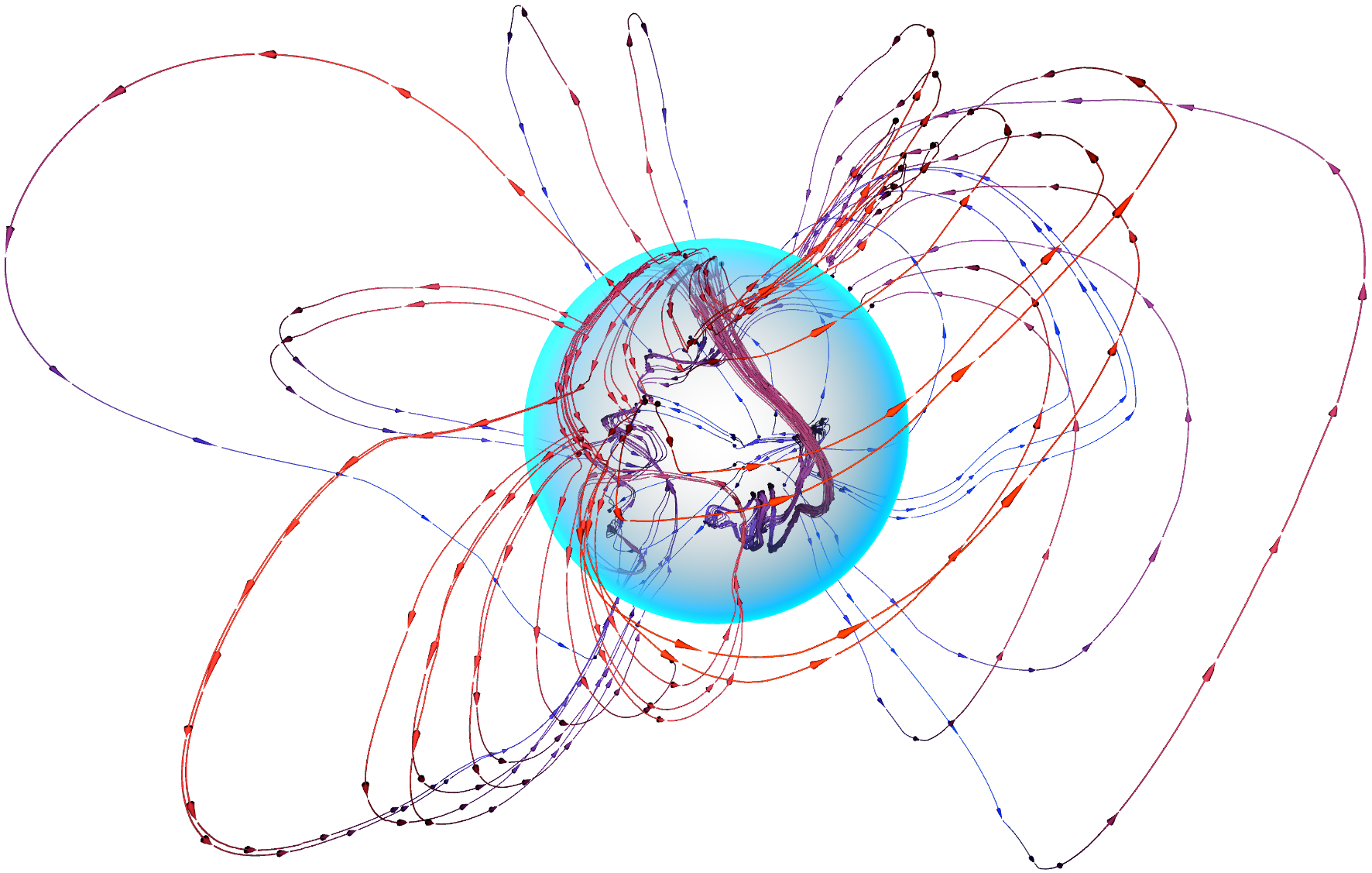}
  \centering
   \caption{Screenshot from the tool. The magnetic field is visualized with streamline arrows and the blue sphere indicates the neutron star.}
 \label{fig:teaser}
\end{figure*}

In this work, we establish a tool for visualizing the magnetic field development during neutron star simulations.
Current workflows struggle with programming overheads and unhandy use of visualization tools and thus hinder the seamless further development of the simulations.
Our tool is directly tailored to the needs of astrophysicists to discover and explore simulation outputs.
We present various techniques that help astrophysicists to investigate in more detail how simulation parameters can be adjusted, and enables domain experts to explore the simulation outcome.
We attach great importance to the interactivity of the tool without long loading times so that questions can be discussed directly.
The techniques were developed in close cooperation with physicists and finally evaluated for their usability.
For an impression of the magnetic field visualization, see Fig. \ref{fig:teaser}.
We combine a 2D \textit{line integral convolution} (LIC) approach suitable for dense fields with 3D data. 
Since a dense 3D data field is difficult to visualize with LIC, we use an interactive 2D cross-sectional surface to project the 3D vector field.
We use this area as an auxiliary area for setting seed points. 
These seed points are then used to create 3D streamlines. 
In this way, we combine a dense vector field representation with a sparse 3D vector field representation using streamlines.
Other parameters used for spatial orientation, such as the mass of the neutron star, are optionally represented by ray tracing or approximated by a mesh.

In summary, our contributions are:
\begin{itemize}
    \item We provide techniques for exploring time-variant magnetic fields by streaming large data sets in real-time
    \item We enable astrophysicists to gain knowledge and insights into the simulation data
    \item We qualitatively evaluate the developed tool by domain experts
\end{itemize}
Within this work, we describe implementation details and concept adjustments. We employed Unity to achieve high frame rates during interactive use, ensuring a smooth and responsive user experience.
The open-source code is available at:
\url{https://t.ly/wmrWU}.
\section{Related Work}

\textbf{Magnetic Field Simulation}.
Evolutions of magnetic field instabilities in single neutron stars in general relativity have been extensively performed in various configurations with differing assumptions on the physical system.
In \cite{Kiuchi:2008ss}, toroidal fields for magnetised stars are studied in the case of enforced axisymmetry with a dynamically evolving spacetime geometry.
In contrast, in \cite{Ciolfi:2011xa,lasky2011hydromagnetic,sur2022long} the development of the instability in initially poloidal fields are studied without symmetry, but assuming a fixed spacetime geometry, that does not evolve as a result of the matter and magnetic field evolving on top of it. In \cite{Cook:2023bag}, the evolution of instabilities in poloidal magnetic fields is also studied without symmetry but with a fully dynamically evolving spacetime geometry.

\noindent \textbf{Astrophysical Visualization}.
In the current astrophysical visualization research, many tools exist to illustrate specific domain data.
A vast overview is given by Lan \etal \cite{lan2021visualization}.

They are capable of processing large datasets and have a huge set of visualization instruments.
However, handling our specific data in real-time, such as visualizing the simulation results over time and exploring them without computationally expensive processing for every time step and even displaying an animation of the simulation results, is very difficult.
Furthermore, tools specifically targeting visualization of the magnetic field of neutron stars are very rare and have limited features that are necessary for domain experts to explore the data properly.
Magnetic field computation and visualization are key research areas. Cochrane \etal \cite{cochrane2023magnetic} mitigate spacecraft interference in Europa’s ocean measurements, while Xiong \etal \cite{Xiong2024} visualize terrestrial magnetic field topology in global MHD simulations to improve magnetospheric understanding.
Visualization tools that are frequently used are ParaView \cite{ahrens200536}, VisIt \cite{weber2010recent,childs2012visit}, Inviwo \cite{inviwo2019} (scientific visualization), and Medvis (medical visualization).

\noindent \textbf{Dense Vector Field Visualization}.
Dense vector field visualization can often be used to get an overview of a vector field. 
At the same time, it shows the entire field without gaps so that small but important structures are not overlooked. 
Dense vector field visualizations are often realized by using 2D textures \cite{yusoff2016flow}. 
The most common method for visualizing dense vector fields is the Line Integral Convolution method by Cabral and Leedom \cite{cabral1993imaging}.
Since then, LIC has been improved extensively for other areas like surface rendering, unsteady flow \cite{shen1997uflic,ebert2002auflic,liu2005accelerated} and even terrain generation \cite{geisthovel2018automated,jenny2021terrain}\cite{shen1997using,matvienko2012metric}.
There are various approaches to using LIC in 3D space. To use LIC in 3D space, it is often applied to 3D surfaces \cite{Xiaoyang97,teitzel1997line,lawonn2014line}.
Several authors \cite{689664,663912,falk2008output,Netzel2013TextureBasedFV} extended the 2D LIC method directly to 3D space.
The idea is straightforward: in principle, the calculation of LIC is not limited by the dimension, which is why 3D textures can be used instead of 2D textures for a small adjustment.
However, the resulting 3D image suffers from occlusion and must be displayed using volumetric rendering. 
Therefore, the actually advantageous overview of the dense vector field visualization is lost.
Nevertheless, methods exist to visualize dense fields, for example Park \etal \cite{park2005dense} uses multidimensional transfer functions to highlight the fine structures by means of direct volume rendering. Later, Park \etal \cite{Park2006} used cell based particle tracing to compute dense vector field visualizations.

\noindent \textbf{Streamlines}.
3D streamlines are widely used to visualize 3D flow fields \cite{mcloughlin2010over,Meuschke:2017,yusoff2016flow,liu2020flow,eulzer2021visualizing}. Although streamlines can also suffer from occlusion, optimization methods exist to tackle this issue. Günther \etal \cite{gunther2013opacity} establish a view-dependent opacity regulation to increase the visibility of obscuring streamlines. 
Lawonn \etal \cite{lawonn2014coherent} improved the visibility of streamlines by using suggestive contours, a technique used in illustrative visualization \cite{lawonn2018CGF}, while emphasizing flow characteristics.
In terms of streamline rendering, Han \etal \cite{Han2019} and Kanzler \etal \cite{Kanzler2019} use ray tracing, both to increase performance and transparency correctness.
Everts \etal \cite{Everts2009} use halos to emphasize streamline bundles.
Jobard \etal \cite{Jobard2000} correlate 2D streamlines to create smooth animations of unsteady flow.
Multiple streamlines can be computed well in parallel, as shown by Pugmire \etal \cite{pugmire2009}.
A crucial criterion of using streamlines beneficially is their placement.
For the placement of streamlines, manual and automatic techniques are used \cite{sane2020survey}. 
Laramee \etal \cite{laramee2004investigating} combined texture-based 2D dense flow field visualization with 3D streamlines. 
They used a seed placement grid to let users manually place seed points.

\section{Stakeholder Analysis}
\label{stakeholder}
To develop the program, we consulted an expert team consisting of seven domain experts from the outset, with specialist experience ranging from around two to over ten years.
In the first meeting, we discussed basic problems that occur in their workflow and could be traced back to visualization problems.
After the first meeting, there was a second round in which we discussed specific features and requirements for the tool. 
These features and requirements were then grouped and prioritized.
Regular updates on interim progress were shared and discussed throughout the development process and
concluded with a qualitative evaluation in which domain experts were able to use the tool themselves and give their assessment of its usability.
In the first round, we analyzed the following visualization problems: 
\begin{itemize}
    \item The basic handling of currently used tools for the visualization of astrophysical data, especially magnetic fields, is cumbersome. 
    They use the Matplotlib and ParaView to visualize their data. 
    Although the tools are suitable for generic data, they are not suitable to render real-time exploration visualizations with small overheads of programming and real-time interaction.
    Therefore, some of them cannot be used or do not deliver the desired results. 
    \item To view the data quickly, 2D slices are usually created. 
    Python scripts are used for this task, which works relatively slowly as the potential of the graphics card is not fully utilized. 
    This means that real-time viewing is not possible.
    \item Visualizations thrive on the ability to set different parameters to highlight specific patterns or properties.
    Due to the low level of interactivity in the current workflow, adjusting parameters is very laborious. 
    The visualization process has to be repeated each time parameters are changed.
\end{itemize}

These problems make it particularly troublesome to discuss difficulties in the simulations in real time. 
As an example, the generation of time dependent visualizations of magnetic field streamlines, while possible with current visualization tools, can prove practically difficult for large data sets. A priori it is not obvious which streamlines should be plotted to investigate features of physical importance and so it is incredibly valuable to be able to interactively visualize the data, exploring different streamlines, and to be able to quickly see the evolution of those streamlines as a function of time, to gain an insight into where the physically important behavior is occurring. Using current visualization tools however, the reading in and 3D rendering of this data can prove prohibitively slow to conduct these interactive investigations of different streamlines at all times. Practically, a decision must be made on which streamlines to focus on based on a small number of tests performed on a subset of time slices. Images of these streamlines can then be generated at all time steps by remotely running the visualization tool, and only then can it be assessed whether the evolution of the selected streamlines has successfully captured the physical behavior which was trying to be visualized. If not, further tests must be performed to locate more illustrative streamlines, and the process repeated.

In addition to the problems described, the physicists also noted that they would also like to use visually appealing results for science communication.
From this point, we have derived four main requirements that the development of the tool should adhere to:
\begin{description}
\item[\textbf{R1}] \textbf{Interactive}: The dense field must be clearly visualized for the user, who can decide where to focus at any time. 
Users should be able to explore individual areas of their choice. 
They should be able to select the time step at any time and adjust the visualization as required so that they can work on their particular task.
Changes in the parameters must be immediately visible.
\item[\textbf{R2}] \textbf{Reactive}: The tool should work in real-time and the individual components must be implemented efficiently. 
Inputs should be processed immediately and interaction with the tool should be possible without waiting times.
The data should be loaded quickly and displayed directly.
\item[\textbf{R3}] \textbf{Intuitive}: The discussion of the magnetic field should be the focus. 
The tool should be easy to use so that all the necessary settings are easily accessible and it should be user friendly in general.
\item[\textbf{R4}] \textbf{Engaging}: The tool should be pleasant to look at so that the information is displayed in an appealing way for a wider audience. 
With a decent appearance, graphics and images generated by the tool may be used for presentations in science communication.
\end{description}

In the second round, the required tasks were discussed and approved by the physicists.
We then derived the following interaction techniques from these overarching requirements:
\begin{itemize}
    \item \textbf{Cross Section:} 
    A cross section through the magnetic field should give an impression of the field. 
    At the same time, it should always be freely movable in space so that any point of the field can be displayed.
    As a 2D surface, it can represent the entire dense magnetic field of the cross section without losing clarity.
    \item \textbf{Streamlines:} In contrast to 2D surfaces, 3D fields are difficult to visualize using dense visualization methods. 
    We therefore use 3D streamlines that visualize the magnetic field at domain-specific interesting locations. 
    Colors of the streamlines can transport further properties of the magnetic field to the user.
    The movement of the streamlines can also be used to visualize strength and direction.
    \item \textbf{Seed Selection:} Areas of interest of the magnetic field can initially be observed with the cross section. 
    The cross-sectional plane should be used to set seed points in order to continue exploring exactly there.
    In this way, streamlines can be set very specifically and interesting areas, such as turbulence, can be highlighted.
    \item \textbf{Time Control:} The time controller allows the selection of any point in time for the simulation data. 
    This data is then loaded and serves as the basis for calculating the streamlines. 
    It must be ensured that the data is loaded quickly so that an animation can be played over time.
    In combination with the seed selection, this makes it possible to visualize the development of significant areas.
\end{itemize}

\begin{figure*}[ht]
\centering
\begin{tabular}{cccc}
  \includegraphics[width=0.22\textwidth]{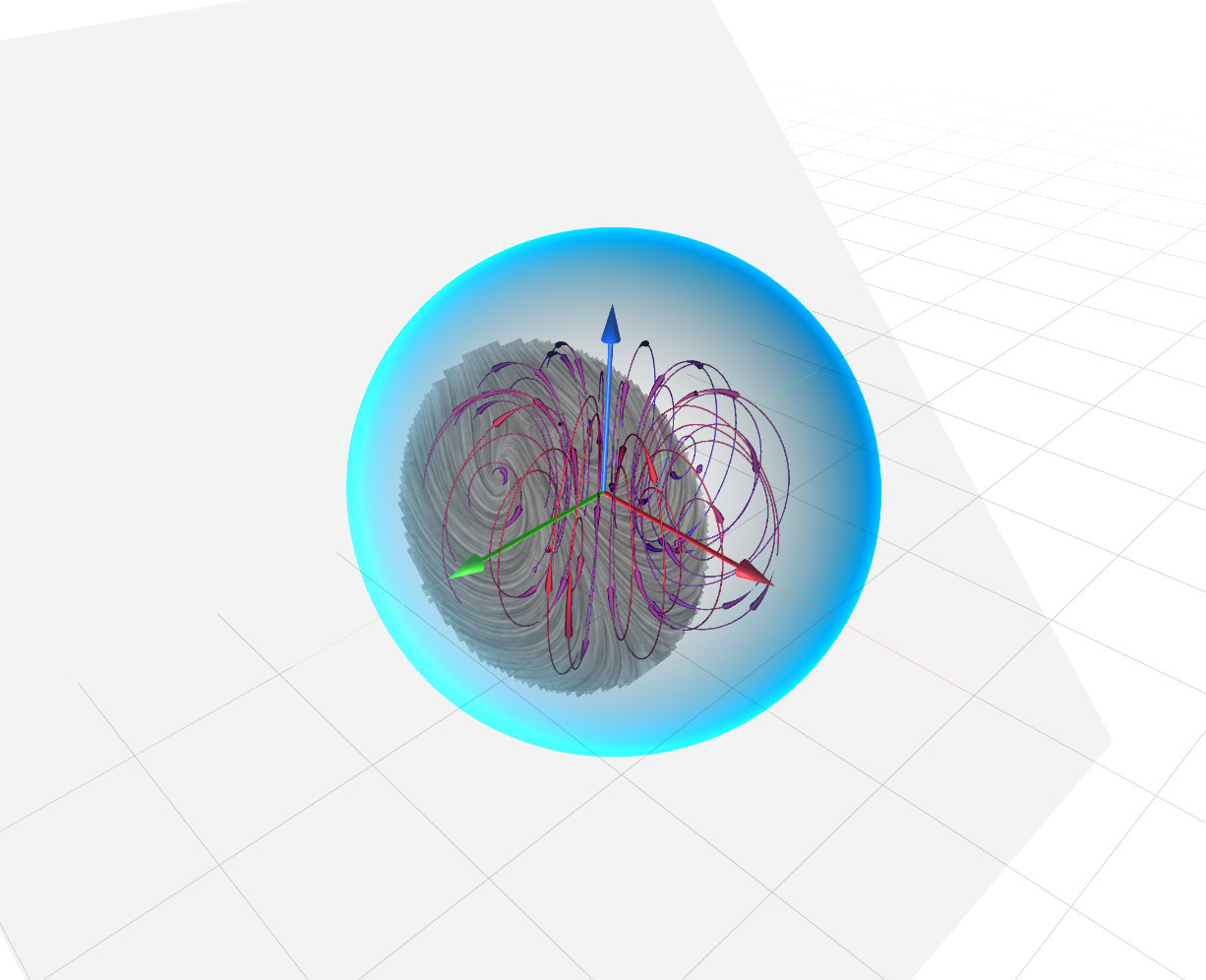}
   & \includegraphics[width=0.22\textwidth]{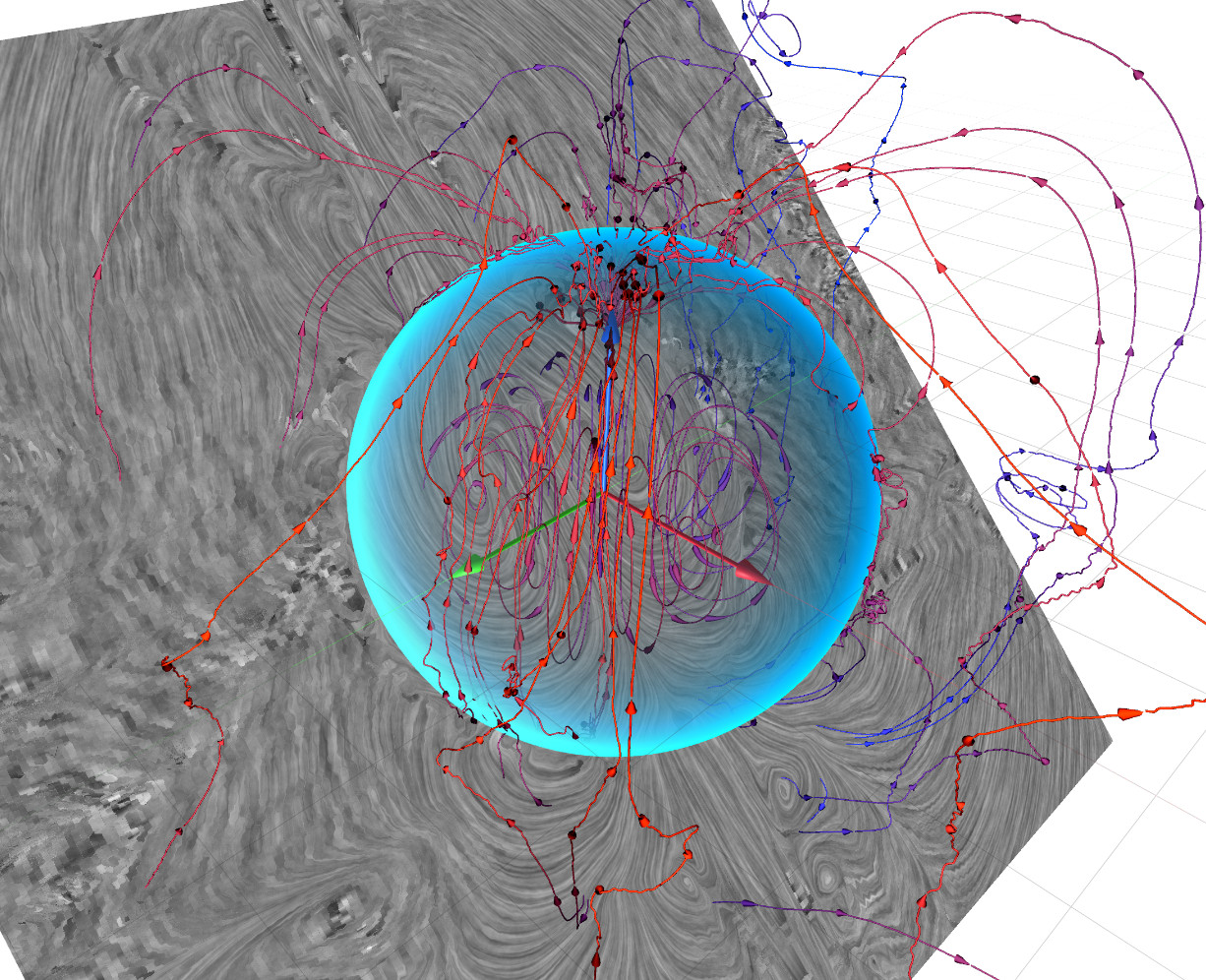}
   & \includegraphics[width=0.22\textwidth]{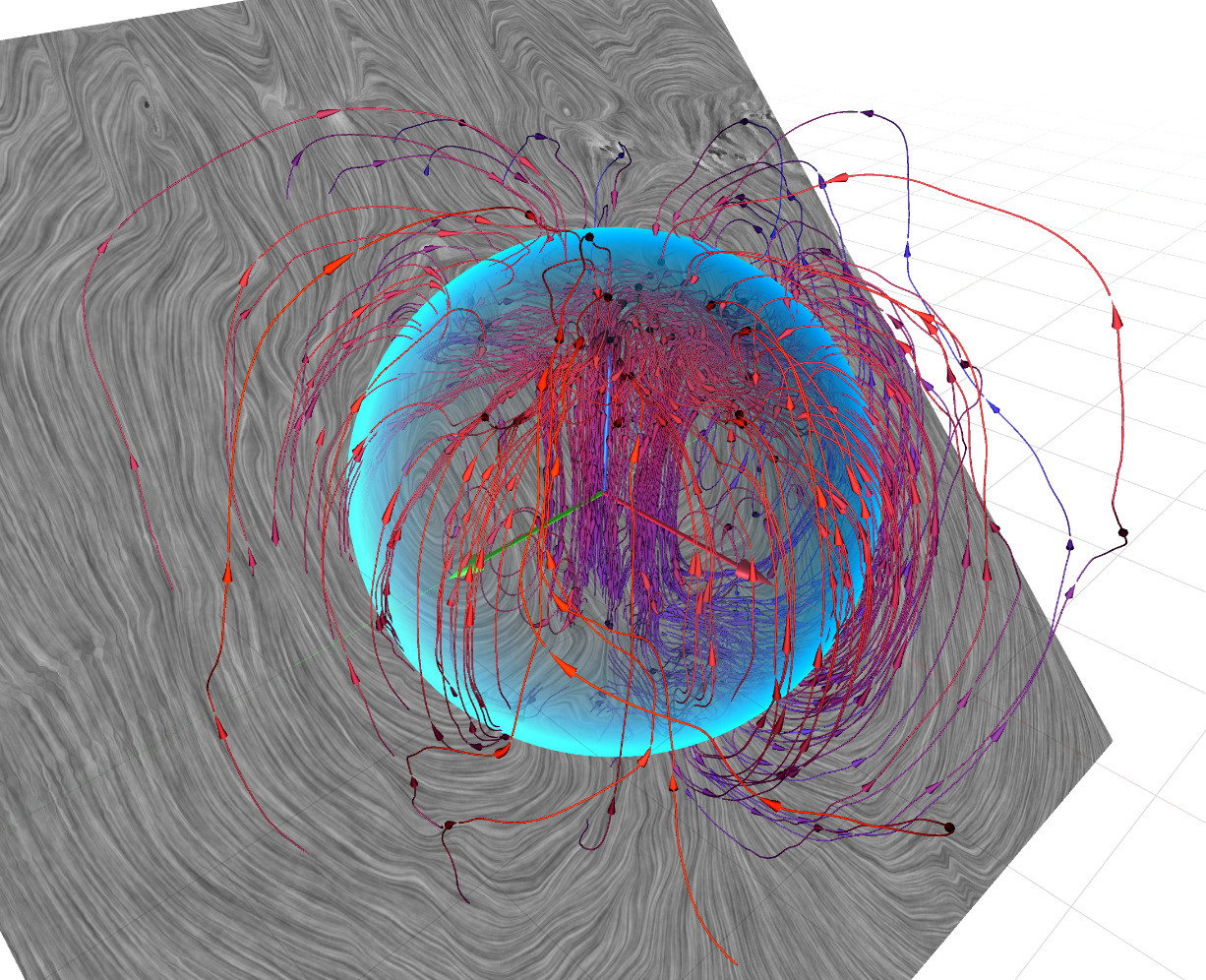}
   & \includegraphics[width=0.22\textwidth]{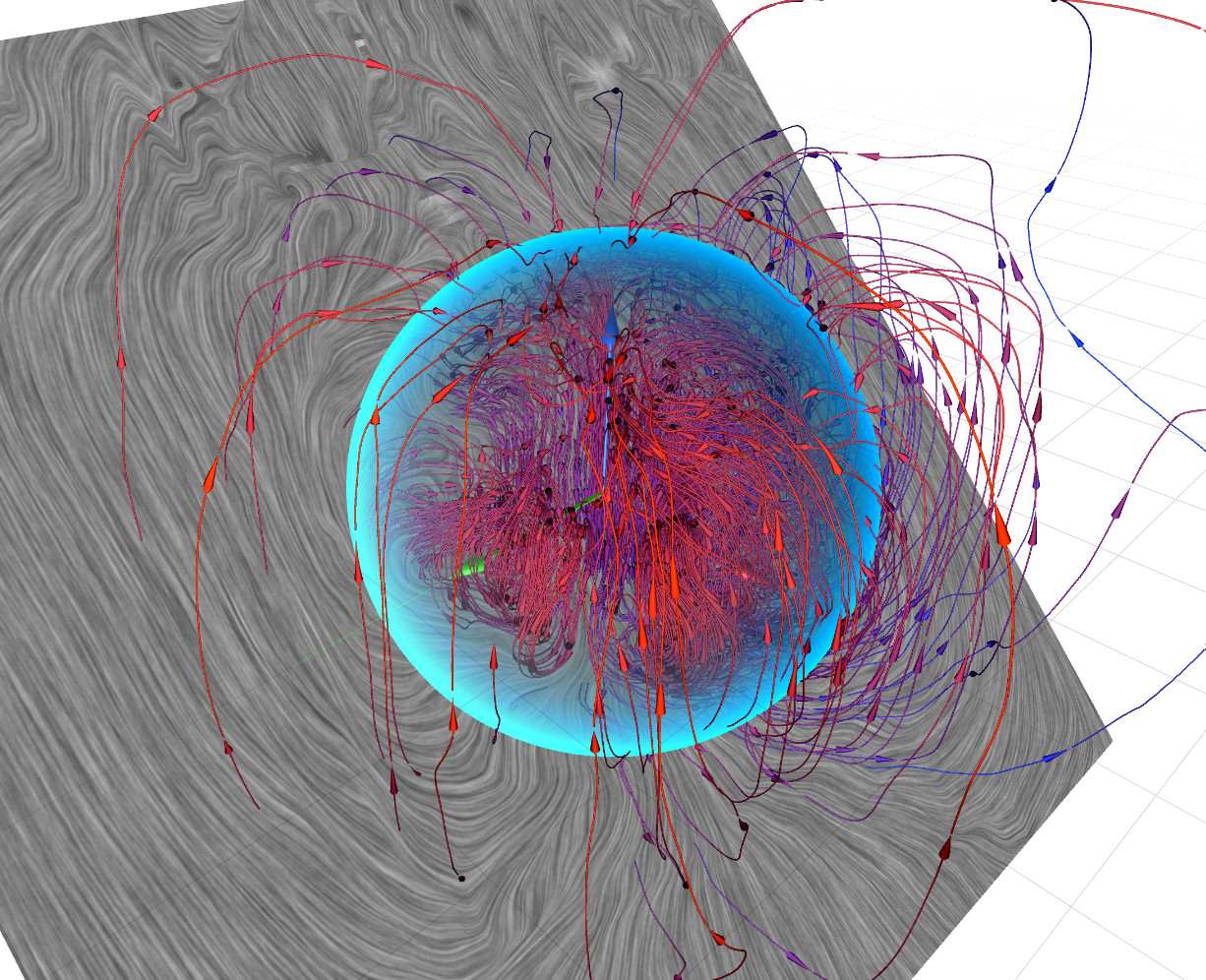}\\
  t = 0 & t = 3  & t = 20 & t = 50\\
  \includegraphics[width=0.22\textwidth]{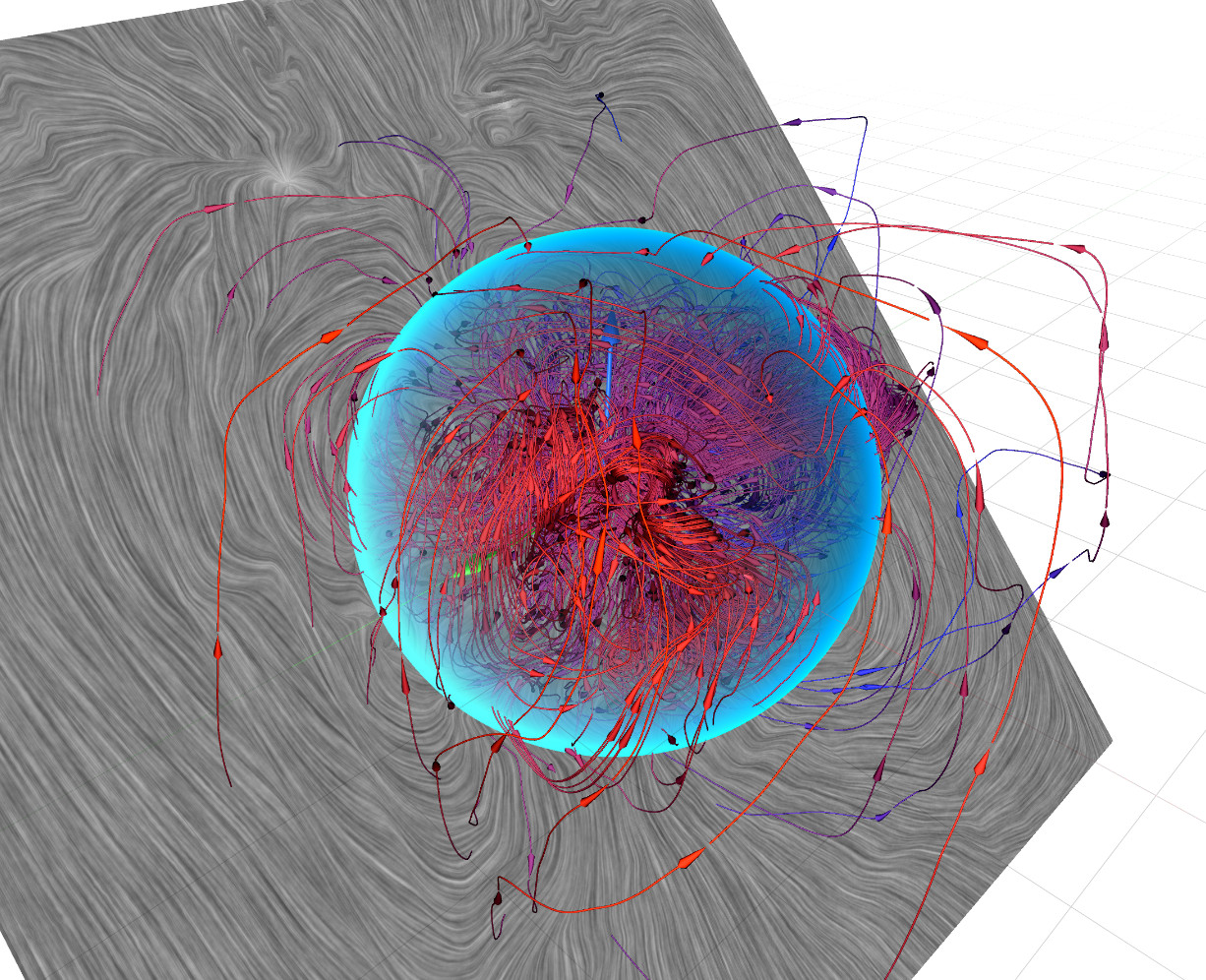}
   & \includegraphics[width=0.22\textwidth]{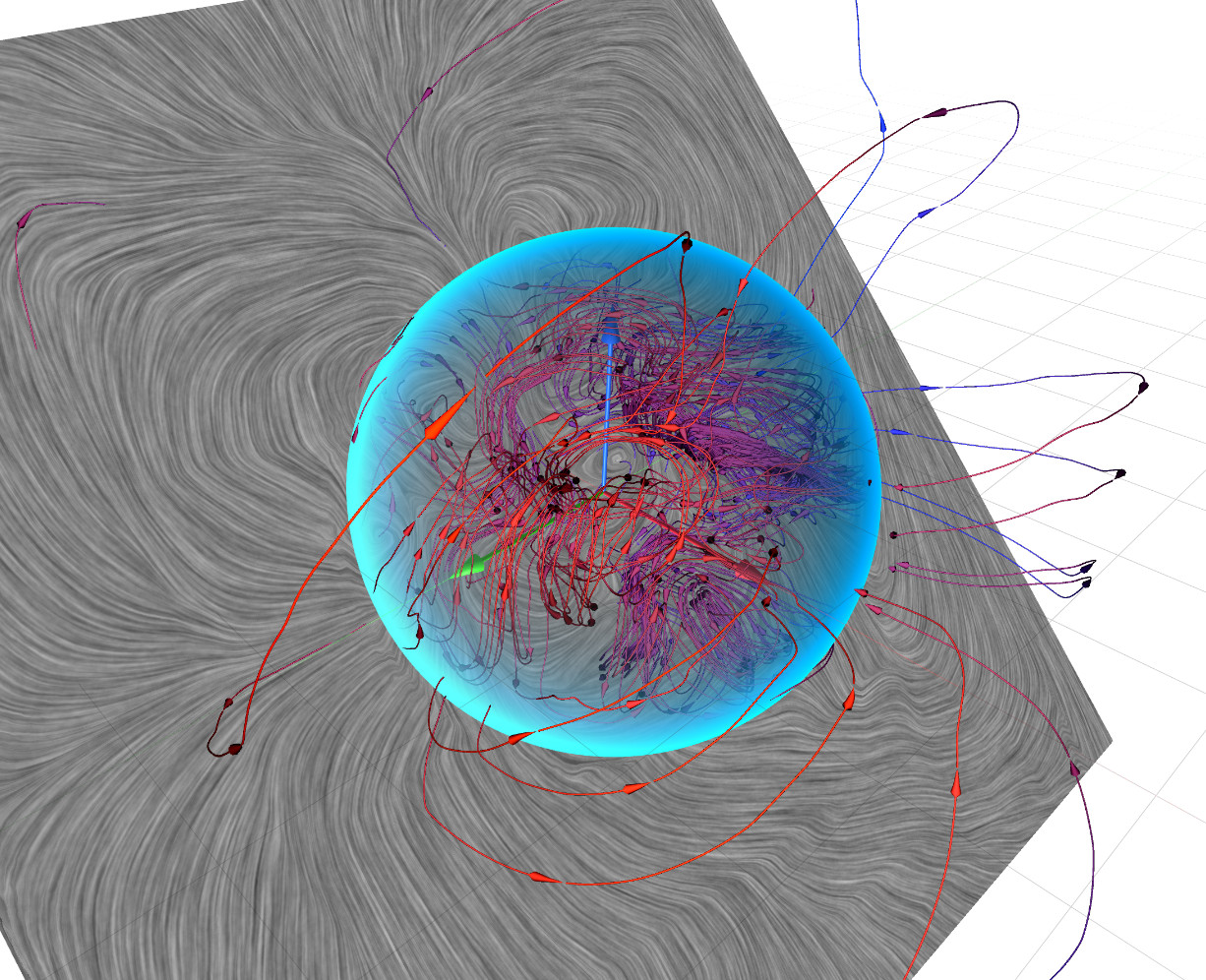}
   & \includegraphics[width=0.22\textwidth]{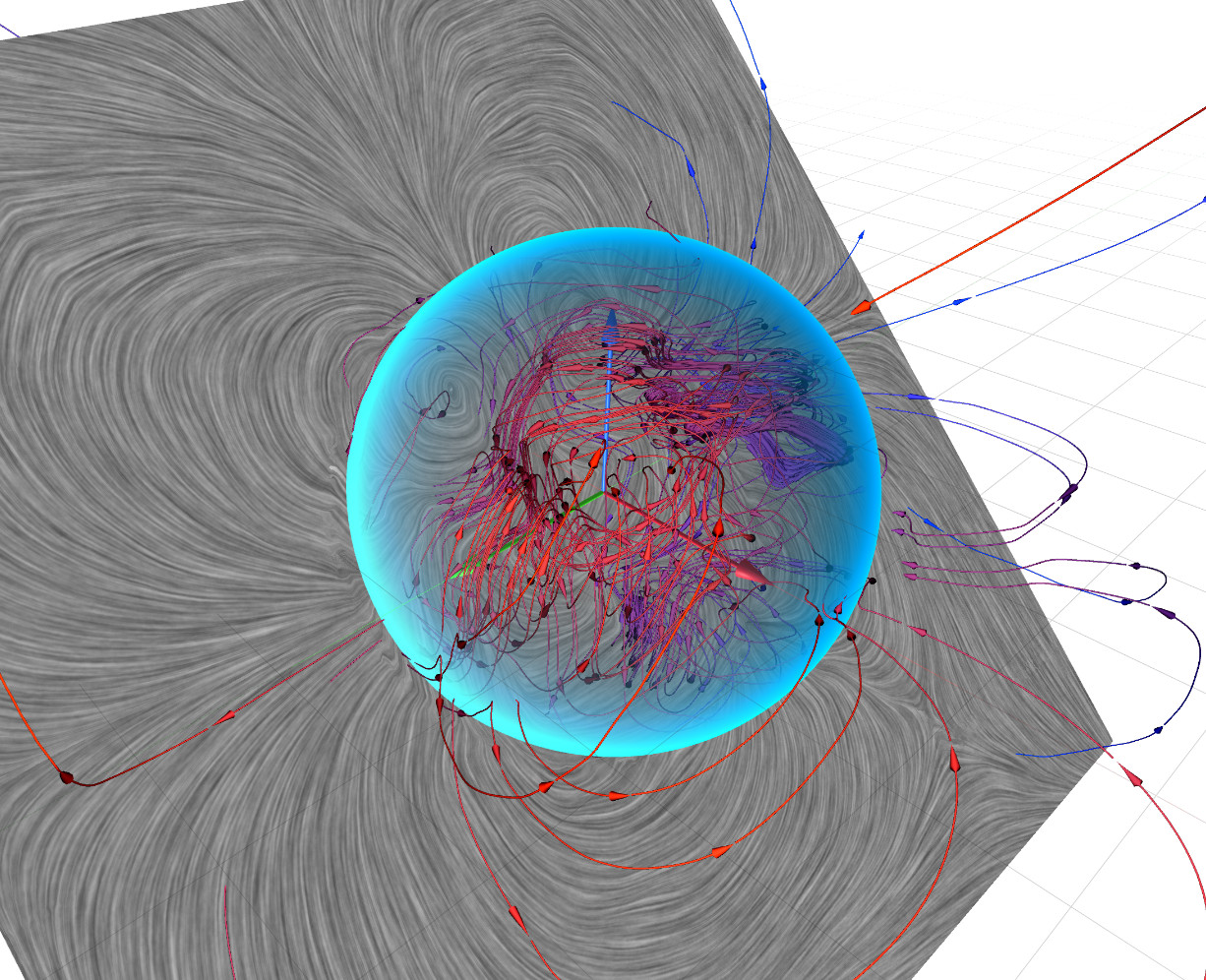}
   & \includegraphics[width=0.22\textwidth]{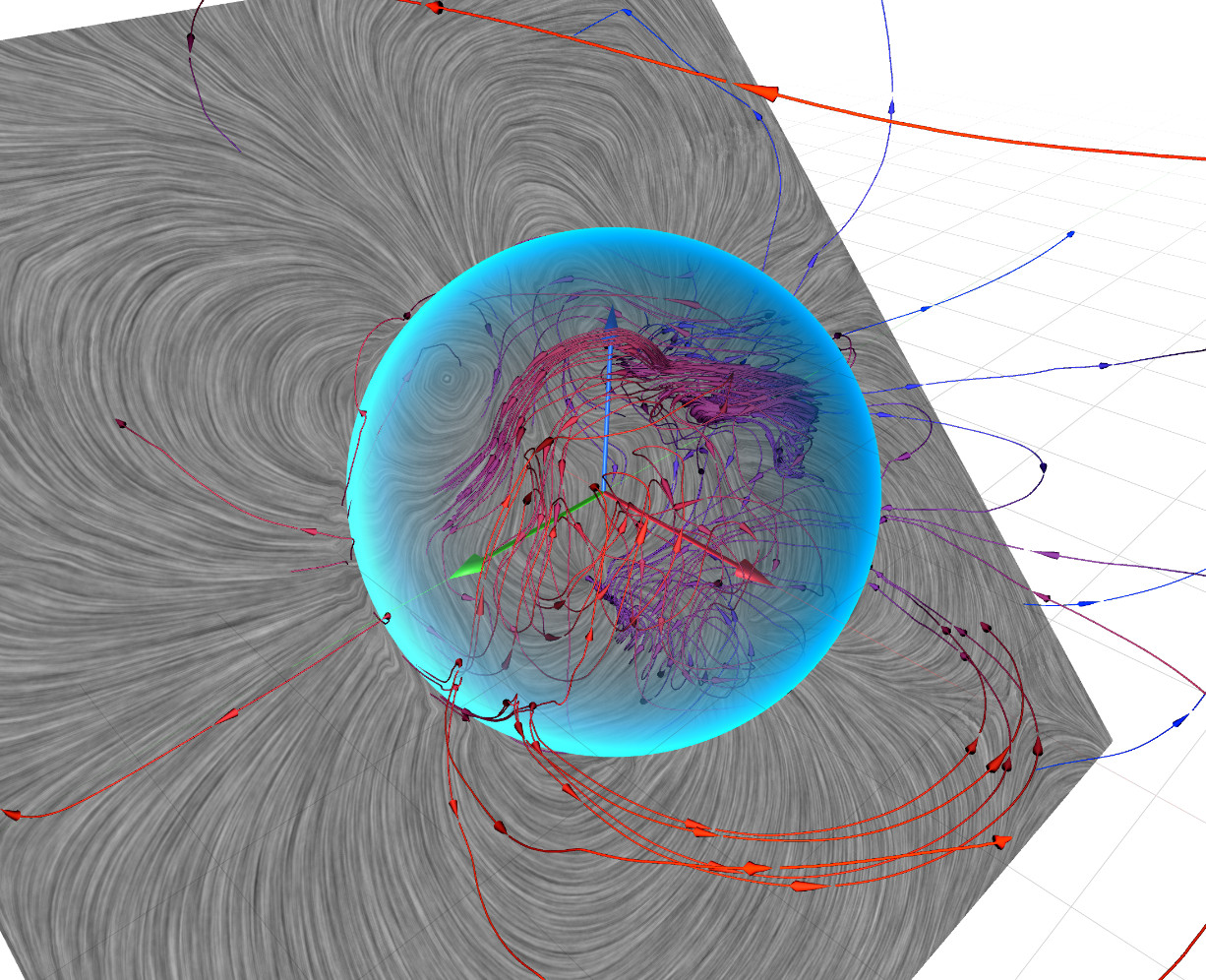}\\
  t = 100 & t = 200  & t = 300 & t = 400\\
  \includegraphics[width=0.22\textwidth]{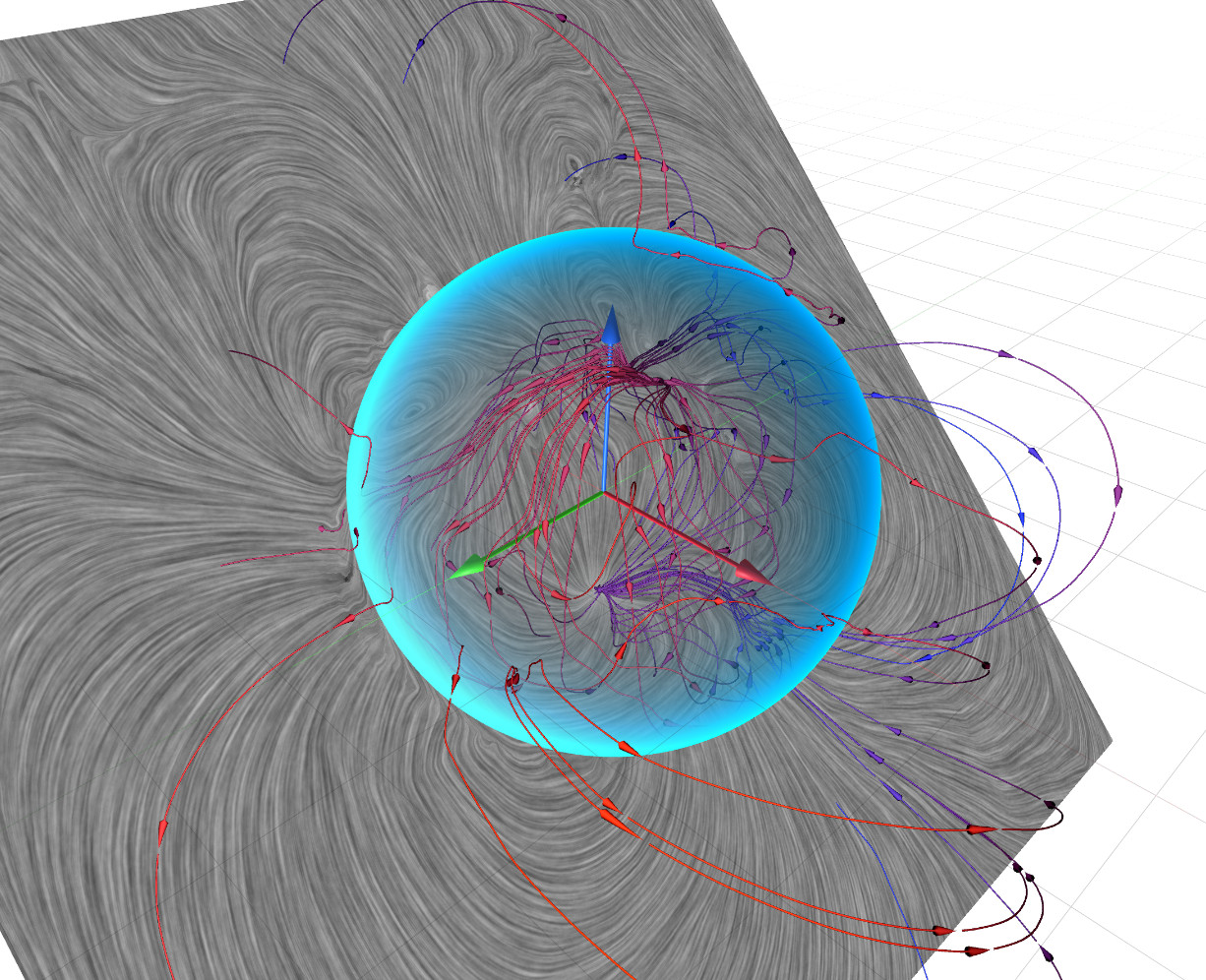}
   & \includegraphics[width=0.22\textwidth]{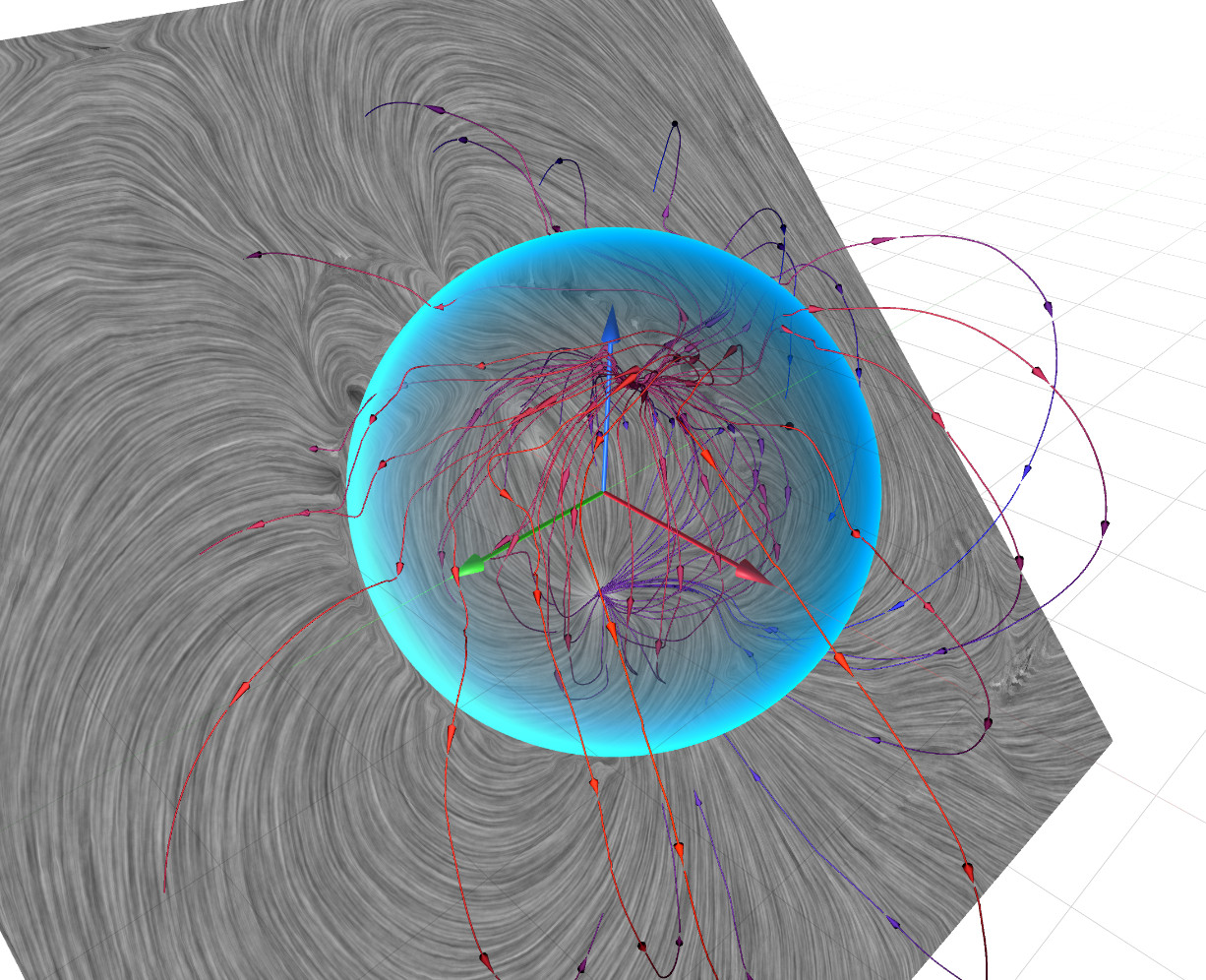}
   & \includegraphics[width=0.22\textwidth]{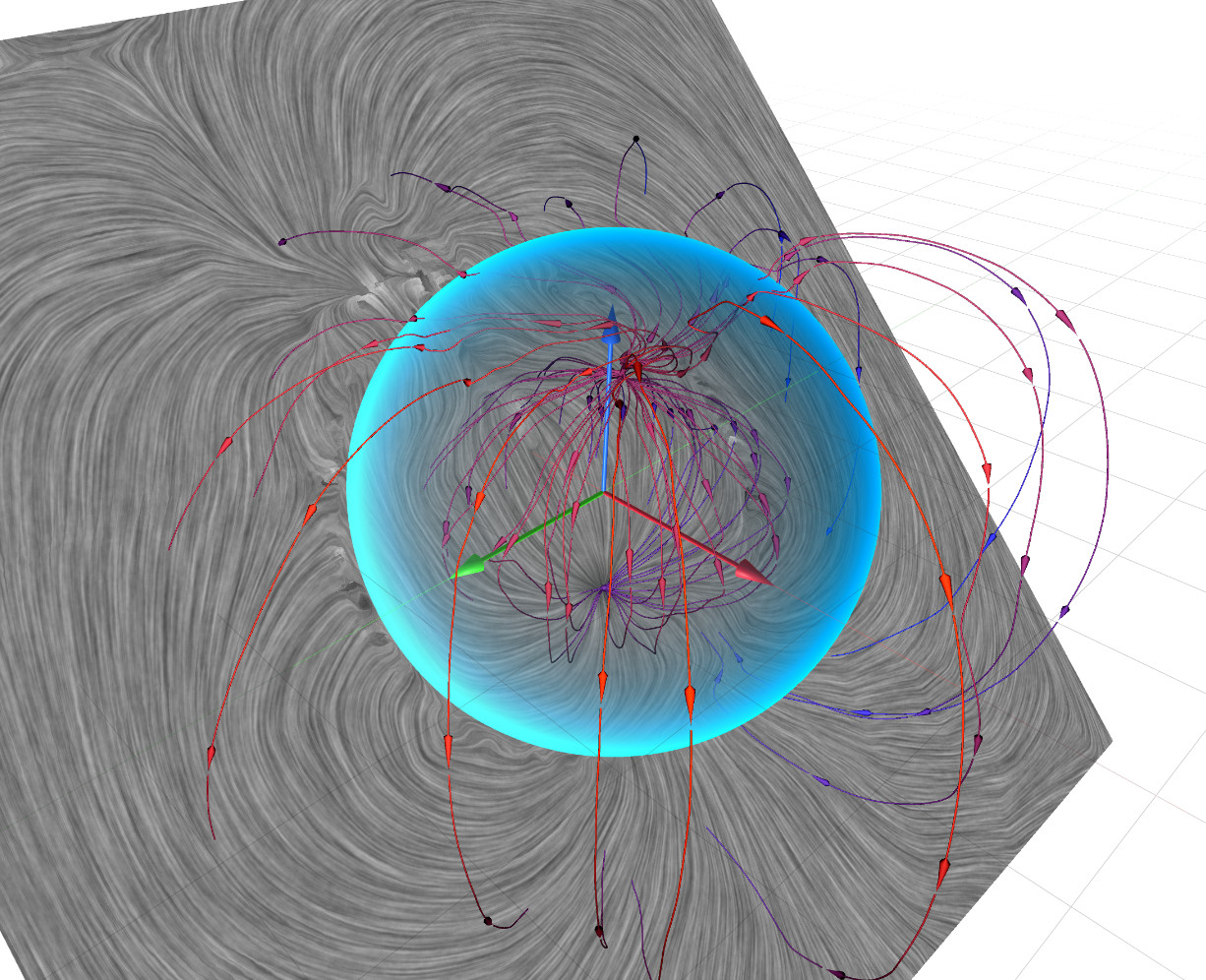}
   & \includegraphics[width=0.22\textwidth]{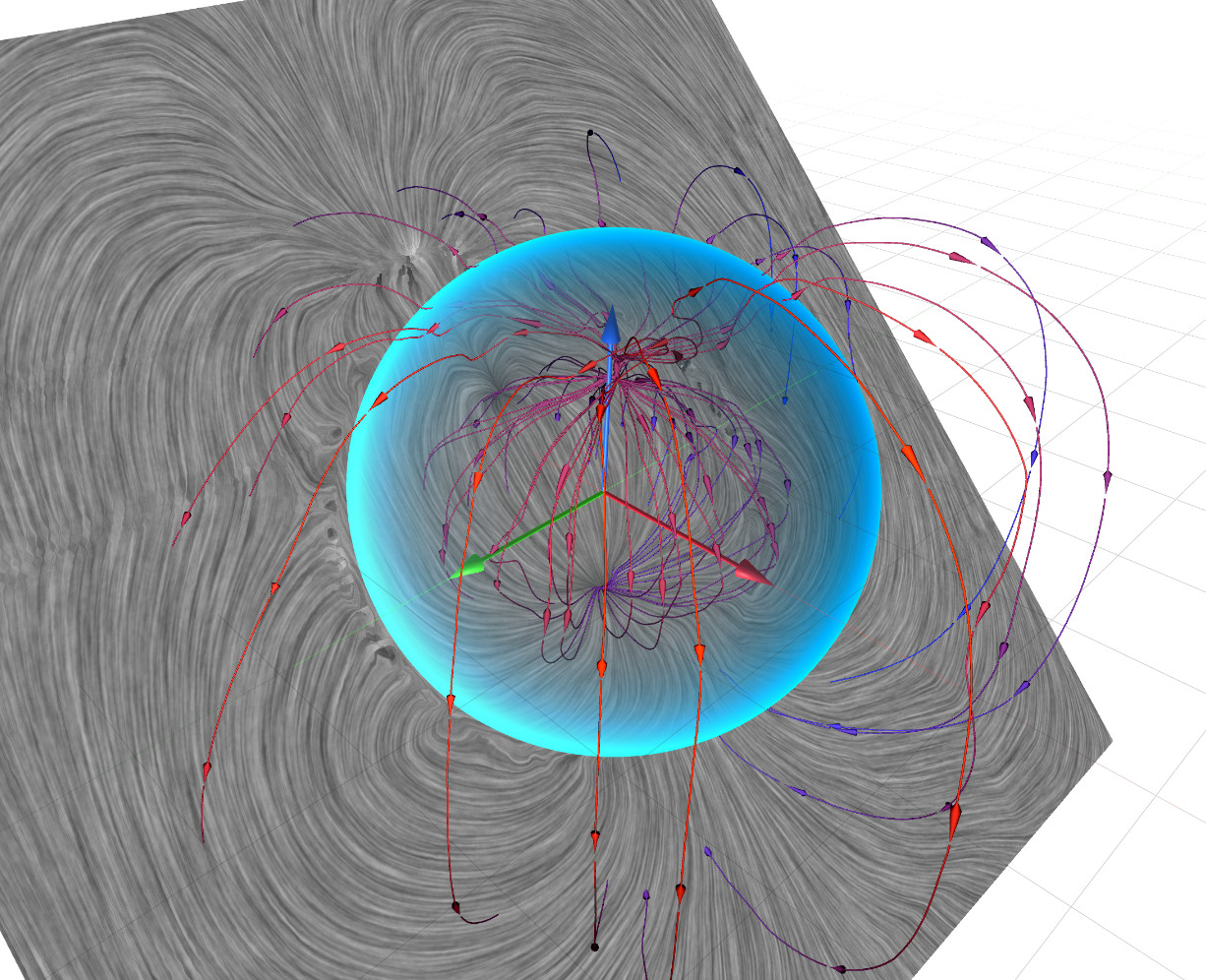}\\
  t = 700 & t = 1000  & t = 1400 & t = 1797\\
\end{tabular}
\caption{The same seed and tool configuration for twelve different time steps (t) 0, 3, 20, 50, 100, 200, 300, 400, 700, 1000, 1400, 1797 (last time step in the data set). The initial magnetic field is set as a toroidal field by default. In the first few time steps, the magnetic field initially spreads out and a lot of chaotic turbulence forms. After around 700 time steps, the field comes to rest and changes only slightly.}
\label{Fig:Evolution}
\end{figure*}

\section{Methods}

In the following subsections, we provide an overview of the implemented features. 
We discuss the methods used and explain their application in the tool.
Finally, we describe a possible workflow with the application.

\subsection{Data Structure, Data Preprocessing \& Data Loading}

The output of data from the simulations is generally given as volumetric datasets in the HDF4 file format \cite{folk2011overview}. 
Each file is approximately 1 GB in size.
With approximately 1,800 files, the total size amounts to 1.8 TB, posing significant challenges for real-time processing during exploration on standard end-user devices.
Since the tool is intended to be used on simple devices such as laptops or desktop PCs, we have derived a pre-processing step from this issue in order to be able to stream the data efficiently.
Each HDF4 file is structured as a spatially uniformly resolved data field, arranged in groups of sub-blocks. 
The individual data points can be both scalar values and 3D vectors.
The arrangement of the data as sub-blocks is useful for processing on many CPU cores in parallel.
However, to utilize the HLSL Sampler State to provide fast data sampling, we restructure the data to be accessible as a 3D texture.
We extract the different data sets contained in one file, e.g. the magnetic field and density values, and spread them to different files, to be able to stream only the data the user is currently viewing.
We use a pre-processing script that restructures the data so that one data set can be loaded as a 3D texture.
The script output data structure is an array of floats. Three consecutive elements form the x-, y- and z- values of a point. 
Points are first aligned in the x-direction, then in the y-direction and then in the z-direction. 
Data dimensions are given as meta data.
The vector fields described in this paper have a dimension of 256³ and thus reach sizes of over 200 MB per time stamp.
In order to be able to load the data onto consumer hardware in real time, we simultaneously generate lower-resolution data sets for each data set in the pre-processing step.
We have found that two additional resolutions, 128³ and 64³, are sufficient to guarantee fast loading times.

Later, the data is loaded into RAM asynchronously to the running program. 
We use two buffers: one buffer always holds loaded data, while a second buffer is used to load new data in the background. 
This ensures that streamlines can be calculated at any time with the buffer always loaded. 
The calculation of streamlines and the reloading of new data can, therefore, also run in parallel.
The data is then copied to the V-RAM as 3D textures.
A time slider is used to select the time step of the data.
Users can change the time step at any time. 
It can, therefore, happen that the user selects a different data set during data loading. 
To ensure real-time usage, we use a data management system that manages the data parallel in the background and deletes and loads data as required so that no memory leaks or lags occur.
This dynamic loading of data also ensures that the data is decoupled from the hard drive and RAM. 
At any time, only the visible part of the data is loaded plus the data that is currently being managed by the memory management system. The system, therefore, scales very well with a database of any size.
A visualization of the magnetic field evolution can be seen in Fig. \ref{Fig:Evolution}.

\begin{figure} [t]

    \begin{subfigure}{.5\linewidth}
  \centering
    \includegraphics[width=1\linewidth]{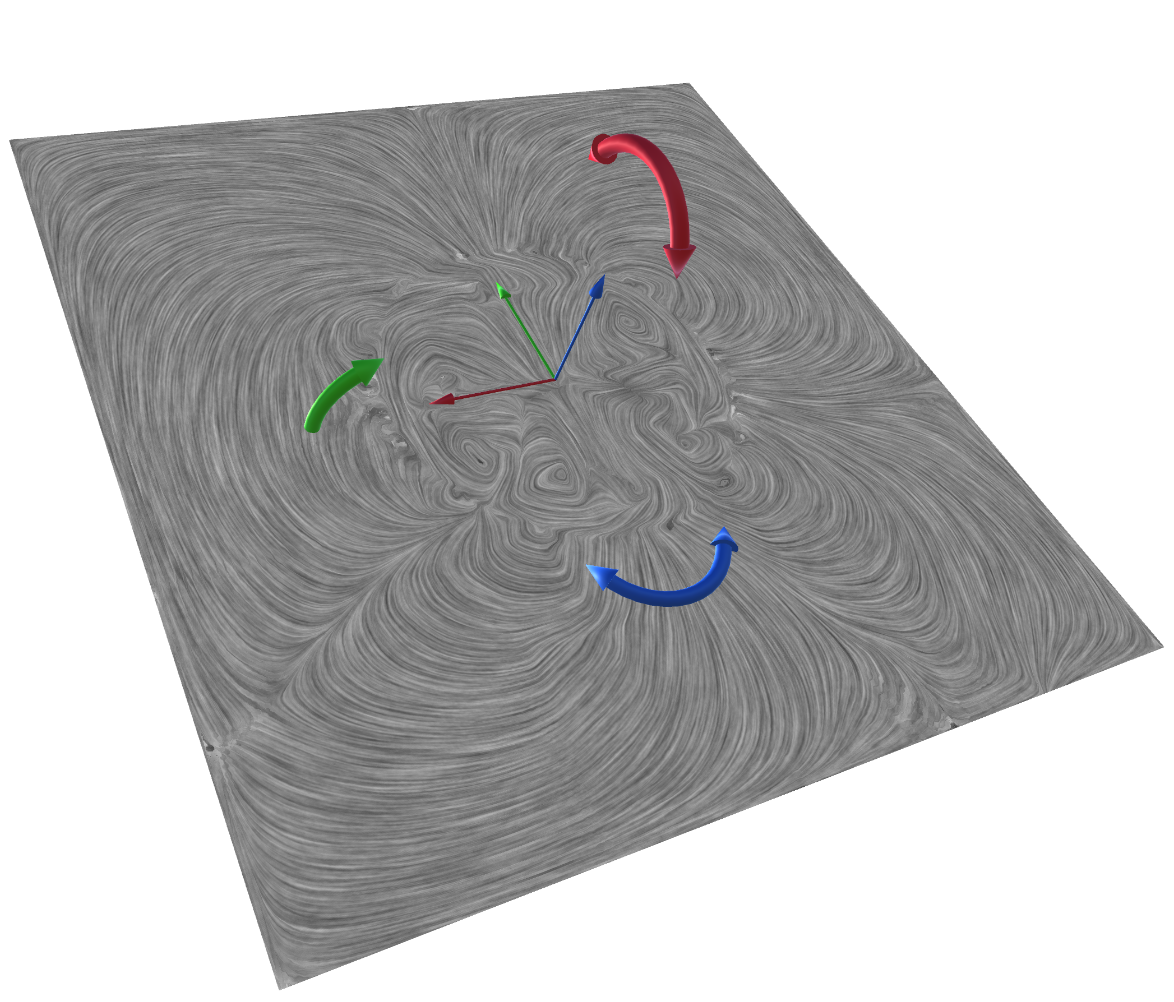}
    \caption{Rotation handle}
  \end{subfigure}%
  \begin{subfigure}{.5\linewidth}
  \centering
    \includegraphics[width=1\linewidth]{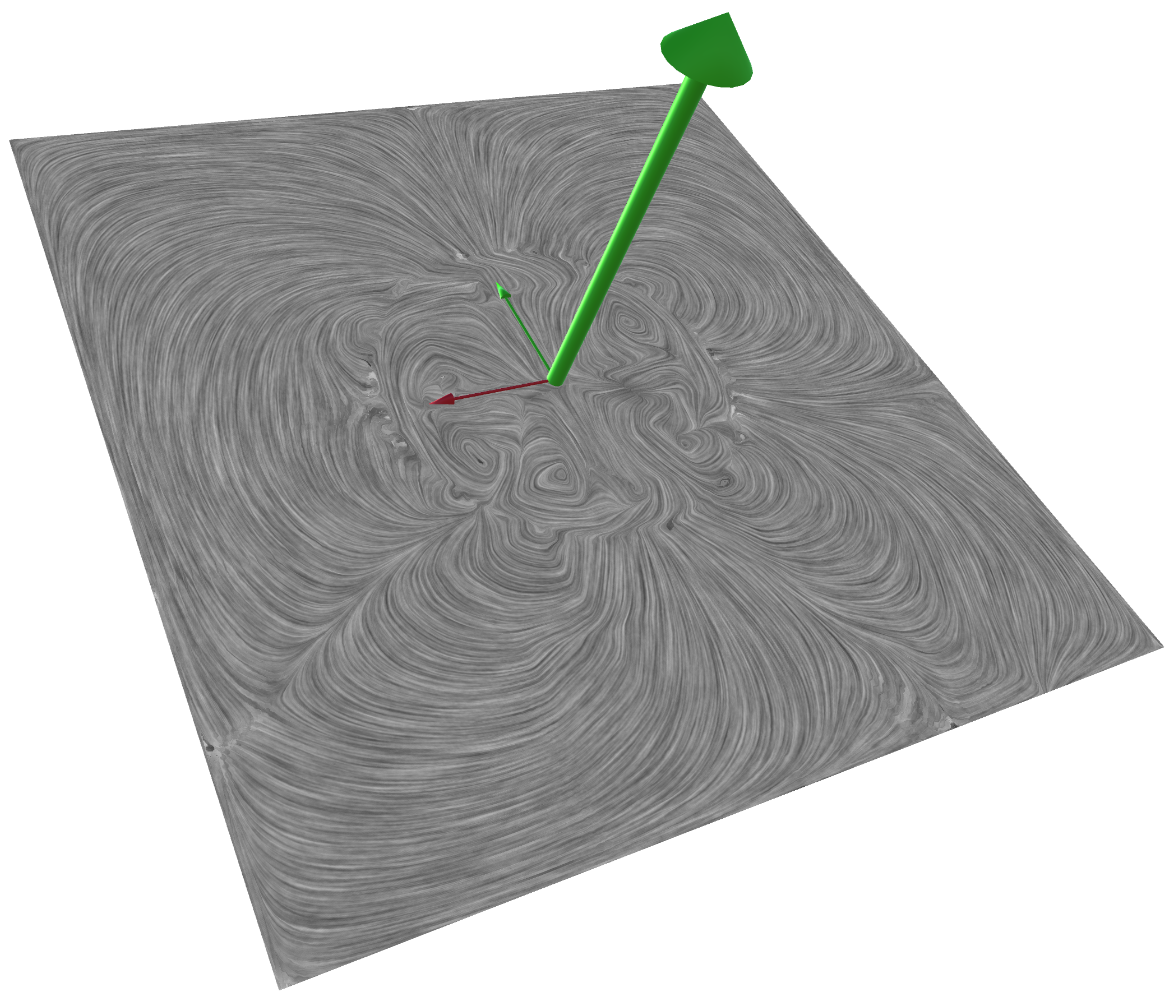}
    \caption{Translation handle}
  \end{subfigure}
 \caption{The two handles control the rotation and position of the Cross Section. a) The rotation handle. b) The translation handle. The Cross Section image is re-calculated in real time.}
 \label{Fig:Handles}
\end{figure}

\begin{figure}[t]
    \centering
    \includegraphics[width=\linewidth ]{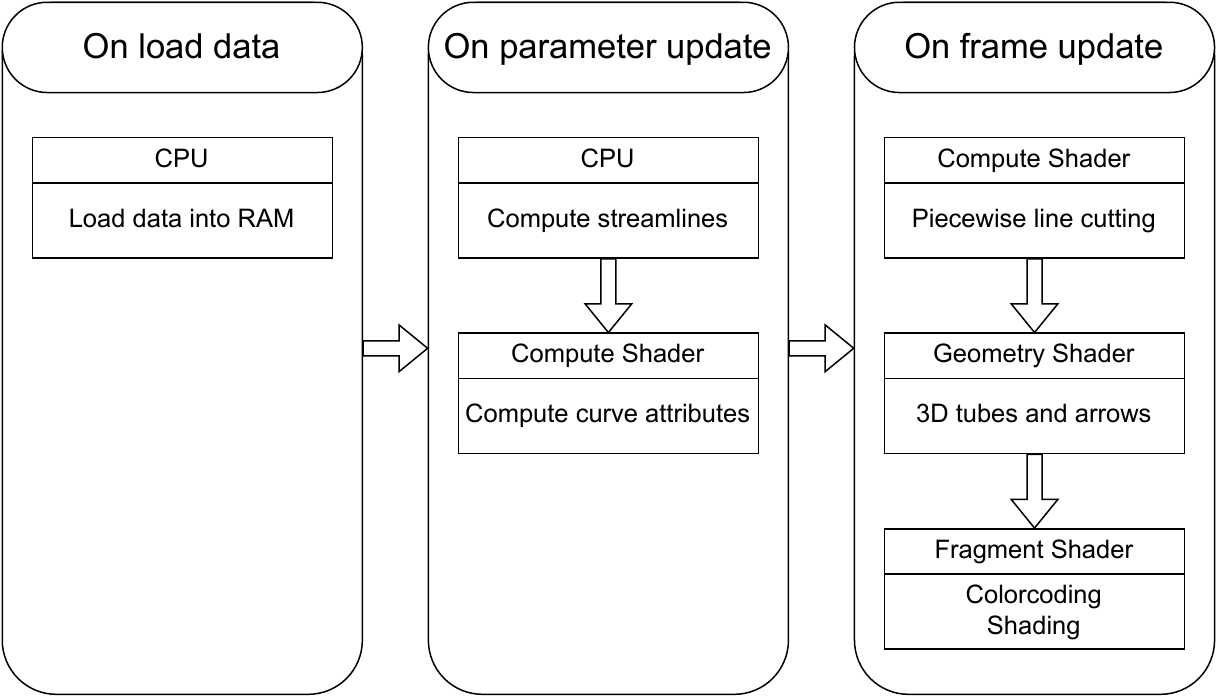}
    \caption{The pipeline to generate and render streamlines. In the \textit{On load data} step, the data is loaded from the disk to RAM. In the \textit{On parameter update} step, the streamlines are calculated in parallel on the CPU using the vector field dataset. Then, the curves are smoothed and attributes are computed in a compute shader. In the \textit{On frame update} step, lines are cut into segments inside a compute shader, and then 3D tubes or arrows are constructed from the streamline skeletons in a geometry shader. In a fragment shader, attributes are color coded and the tubes are shaded to increase 3D perception.}
    \label{fig:streamlinePipeline}
\end{figure}

\begin{figure}[t]
\centering
\begin{tabular}{cccc}
  \includegraphics[width=0.2\linewidth]{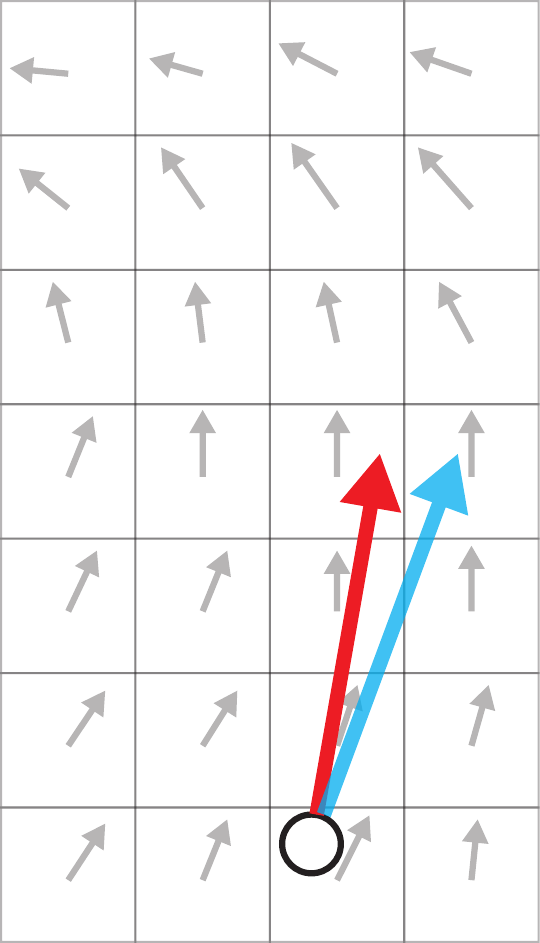}
   & \includegraphics[width=0.2\linewidth]{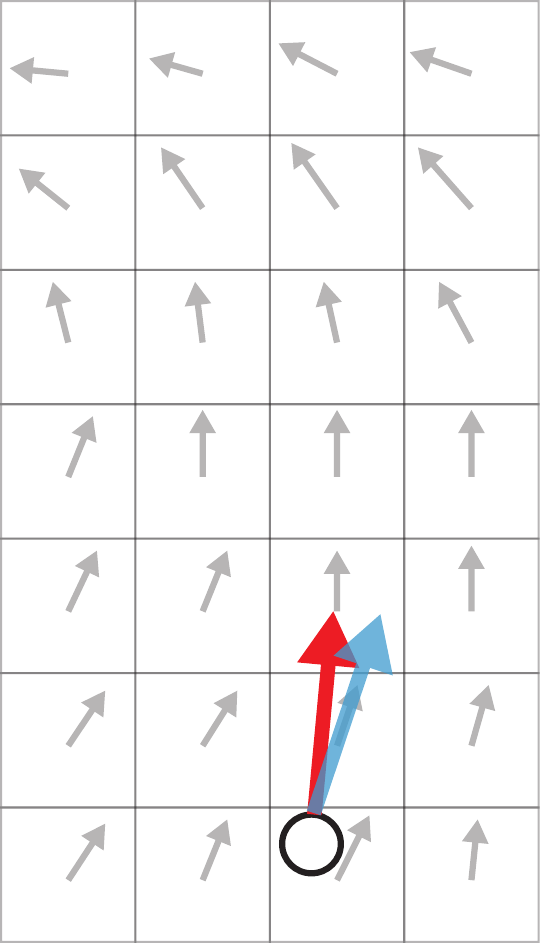}
   & \includegraphics[width=0.2\linewidth]{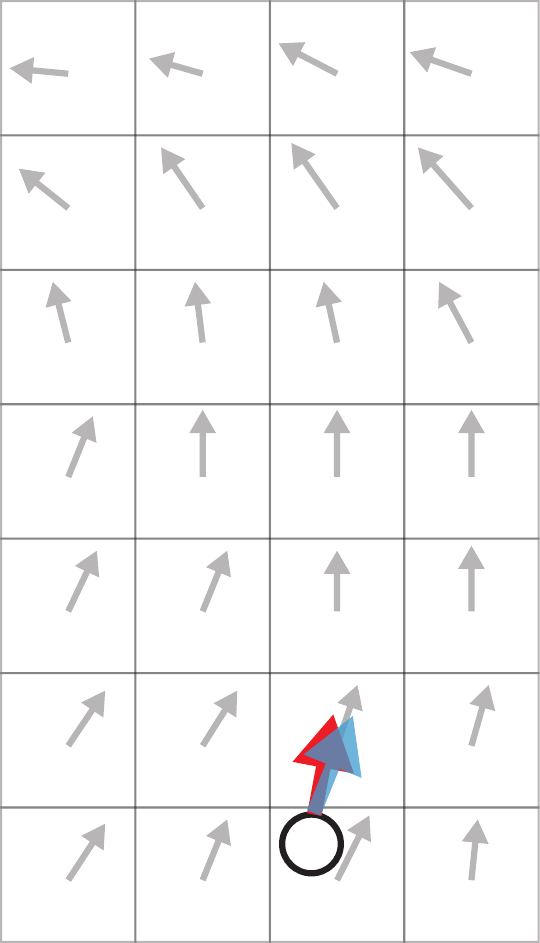}
   & \includegraphics[width=0.2\linewidth]{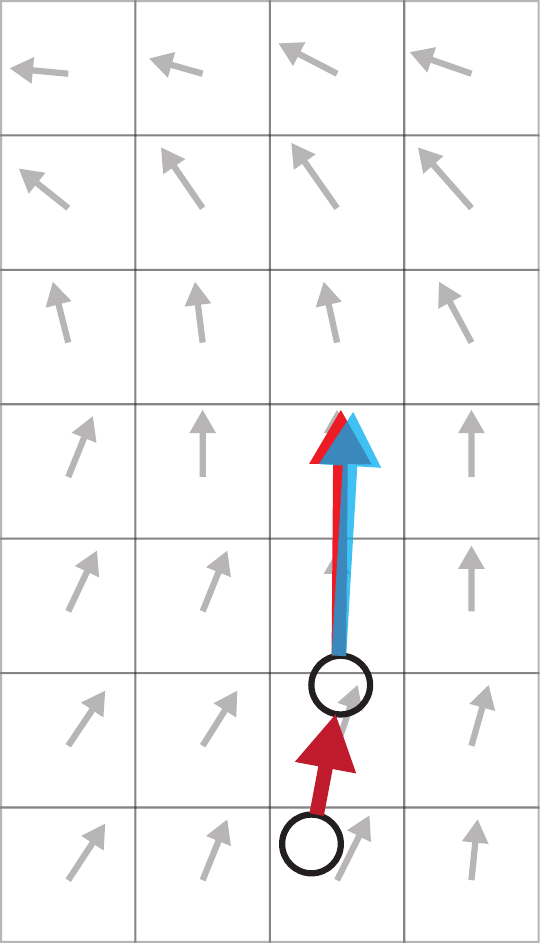}\\
  (a) & (b)  & (c) & (d)\\
  \includegraphics[width=0.2\linewidth]{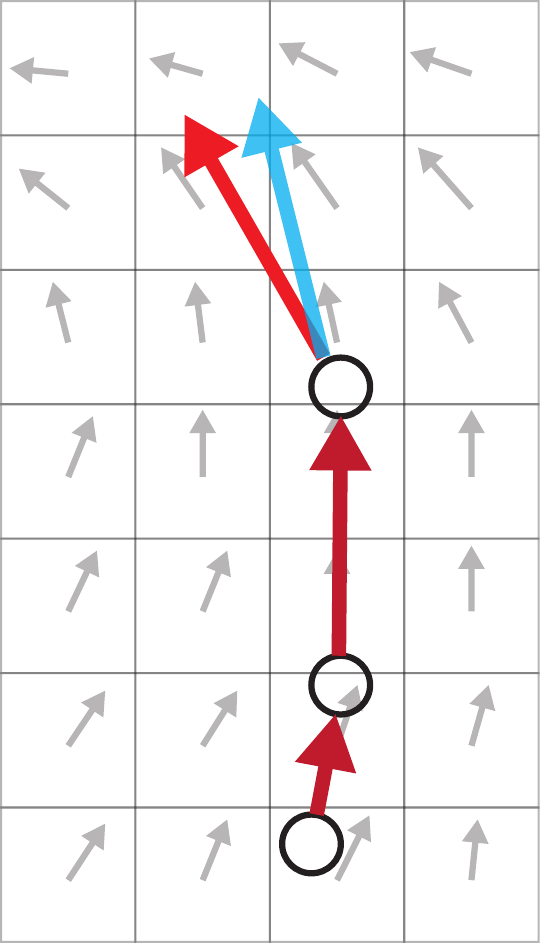}
   & \includegraphics[width=0.2\linewidth]{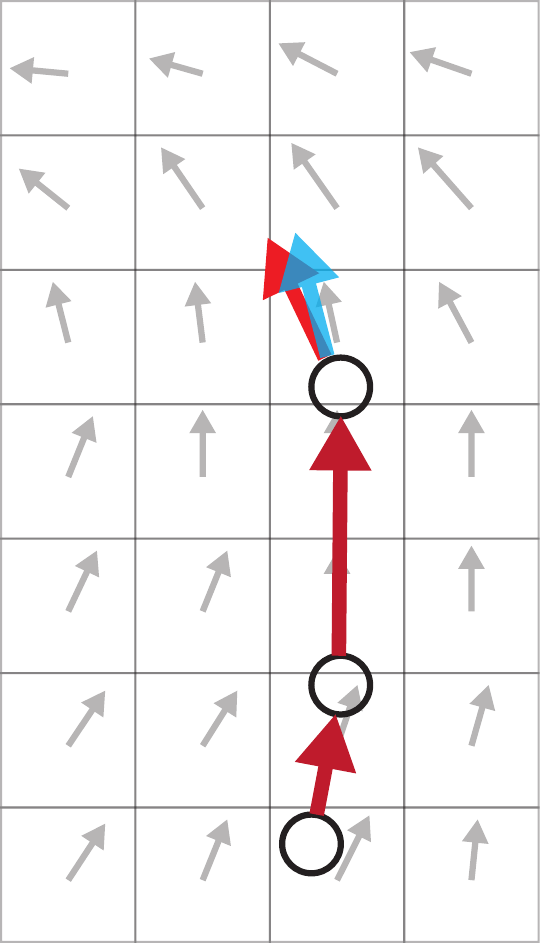}
   & \includegraphics[width=0.2\linewidth]{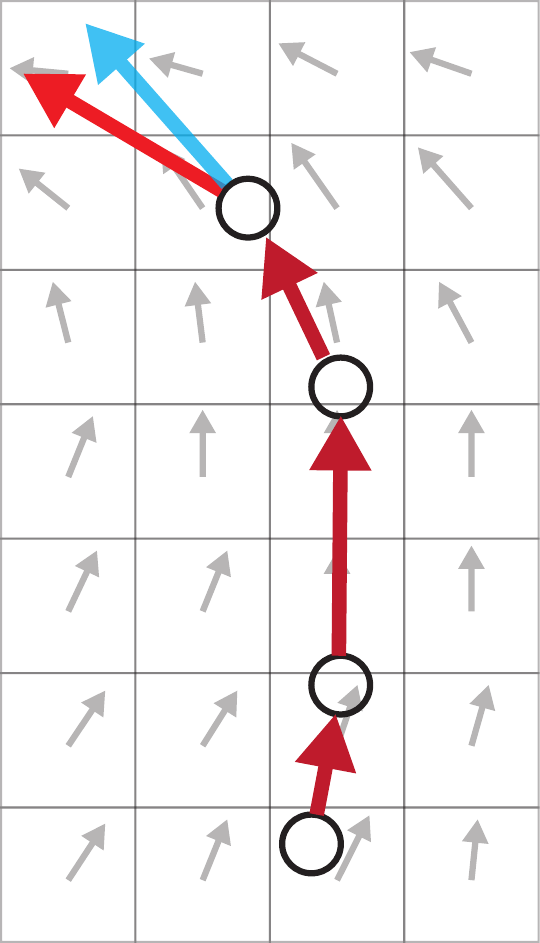}
   & \includegraphics[width=0.2\linewidth]{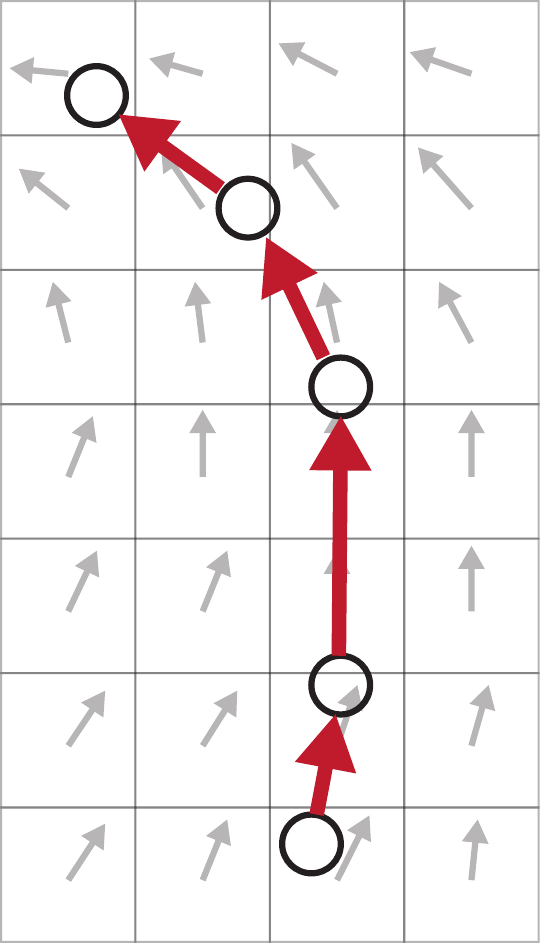}\\
  (e) & (f)  & (g) & (h)\\
\end{tabular}
\caption{Calculation of the streamline using the Runge-Kutta-Fehlberg method. The red arrows indicate the calculation of the next point with 4th order RK. The blue arrows show the calculation of the next point with 5th-order RK. The length of the arrows indicates the stepsize. Set streamline points are indicated by circles. (a) Initial seed point with relatively large stepsize. The difference between the 4th order and the 5th order is very large, so the stepsize in (b) is reduced. (b,c) The error is still too large, so the stepsize is reduced further. (d) The error is small enough so that another streamline point is set and the stepsize is increased again. (f) Now the error is small enough again, as the vector field is quite uniform here. Another streamline point is set and the stepsize is increased. (g,h) This process is repeated.}
\label{Fig:RK}
\end{figure}

\subsection{Magnetic Field Cross Section}

The first interaction technique we implemented is the magnetic field cross section.
It provides an overview of the existing magnetic field and serves as a basis for manual seed point placement.
The plane itself is realized as a quad, which can be moved in space and whose texture has an arbitrary resolution.
LIC is used to visualize the vector field within the plane. 
The real time usage allows to explore areas of interest in the field.
It is realized by two shaders attached to each other. 
In the first shader, a compute shader, the corresponding vector value is determined for each pixel of the texture. 
The spatial position is determined and then the vector is obtained from the data field by linear interpolation between voxel positions.
The vector field is then projected into the plane so that the resulting vectors are 2-dimensional with respect to the plane.

In the second shader, a fragment shader, line integral convolution is applied to the texture.
The shader receives two textures as input: white noise and the vector field, whose values are encoded in the RG channels.
The noise texture serves as a coloring texture. 
For each pixel, its streamline is convoluted in its place. 
This convolution is done on the fly: starting at the position of the pixel, it is integrated into the positive and negative direction using \textit{Runge-Kutta-4} (RK4) \cite{butcher1996history} so that it runs along the streamline. 
The step size here is half a pixel length, but this can also be adjusted.
Users have the option to enable arc-length parametrization with adaptive step size to enhance detail. However, our findings indicate that the improvement in accuracy is minimal, while the associated performance costs are substantial. As a result, we leave the decision to utilize arc-length parametrization up to the user.
The color value of the noise at the calculated position is then calculated in each integration step. 
This color value is weighted according to the step iteration and added to a final value. 
The weighting of the color values is Poisson distributed in regards to the iteration step in order to give greater weight to values close to the point of origin.
As the total error increases as the integration moves further away from the original point, the weighting also reduces the error. 
Overall, the number of steps depends on the step size. 
A step size of half a pixel size and 20 iterations of RK4 integration are sufficient for a clearly visible result.
Since each point is now convoluted along its streamline, points on the same streamline are convoluted with the same line integral shifted by the spatial distance. 
The result is that the noise along the streamlines is "smeared" and the vector field becomes visible.

The cross-section plane is equipped with two handles: It can be moved along its normal and rotated around the X, Y, and Z axes using three-axis controllers, see Fig.~\ref{Fig:Handles}. 
As the shaders are executed in each frame, the current cross section of the magnetic field is always visible. 
This means that users can move the plane interactively through the magnetic field and switch simulation time steps in order to search the field for interesting points.

\begin{figure*}[t]
\centering
\begin{tabular}{cc}
  \includegraphics[width=0.47\textwidth]{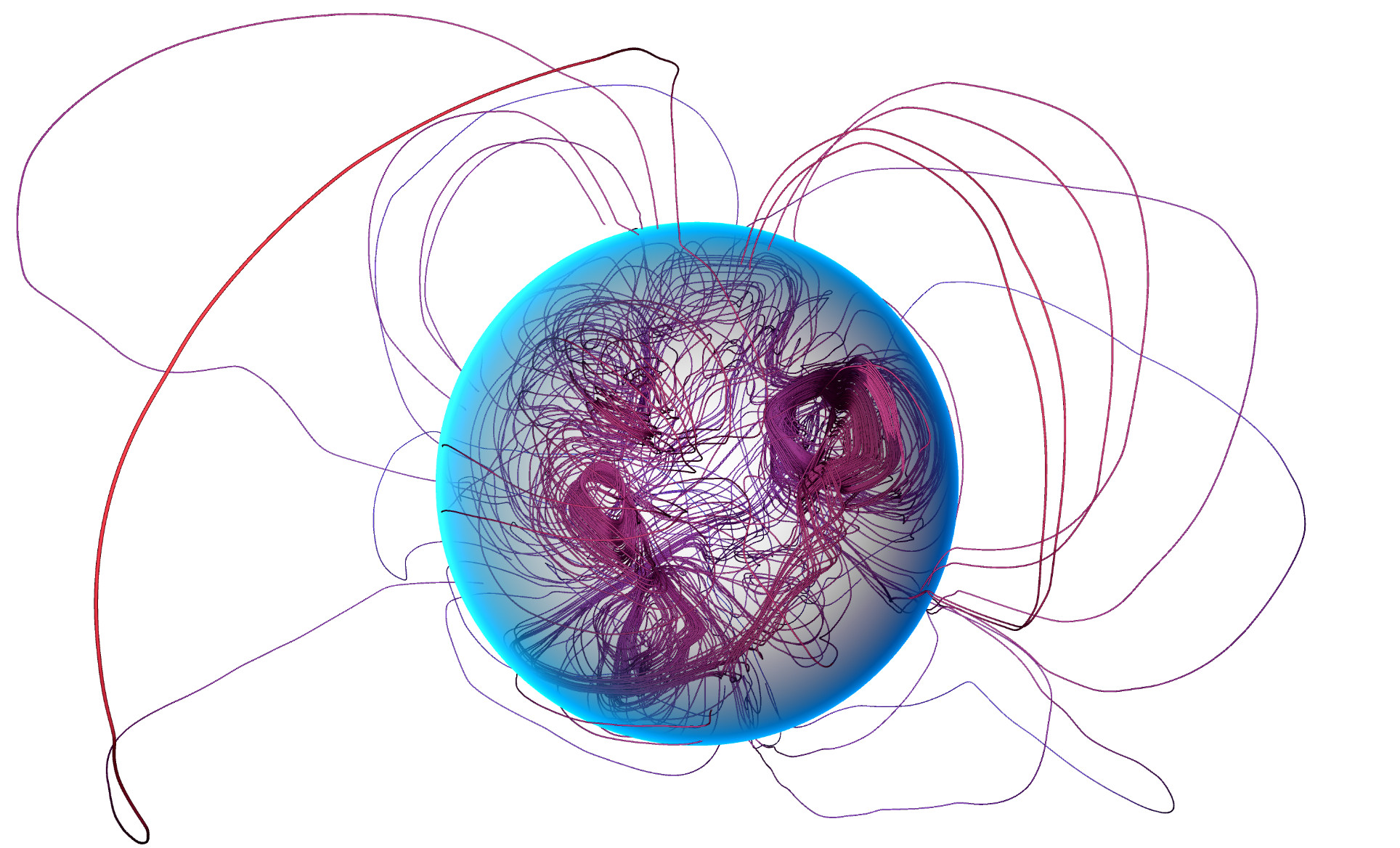}
   & \includegraphics[width=0.47\textwidth]{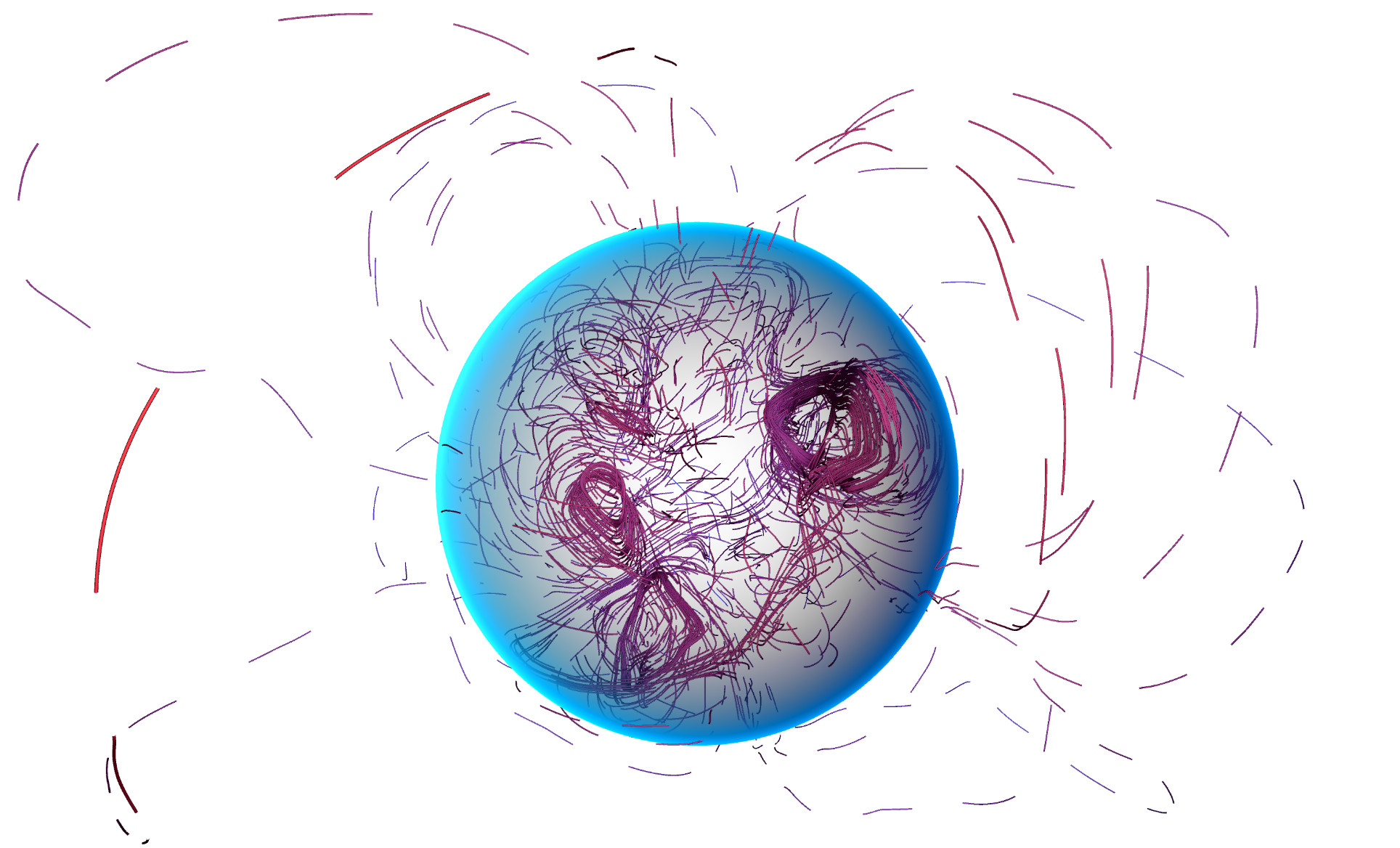}\\
  (a) Full streamlines & (b) Dashed streamlines \\
  \includegraphics[width=0.47\textwidth]{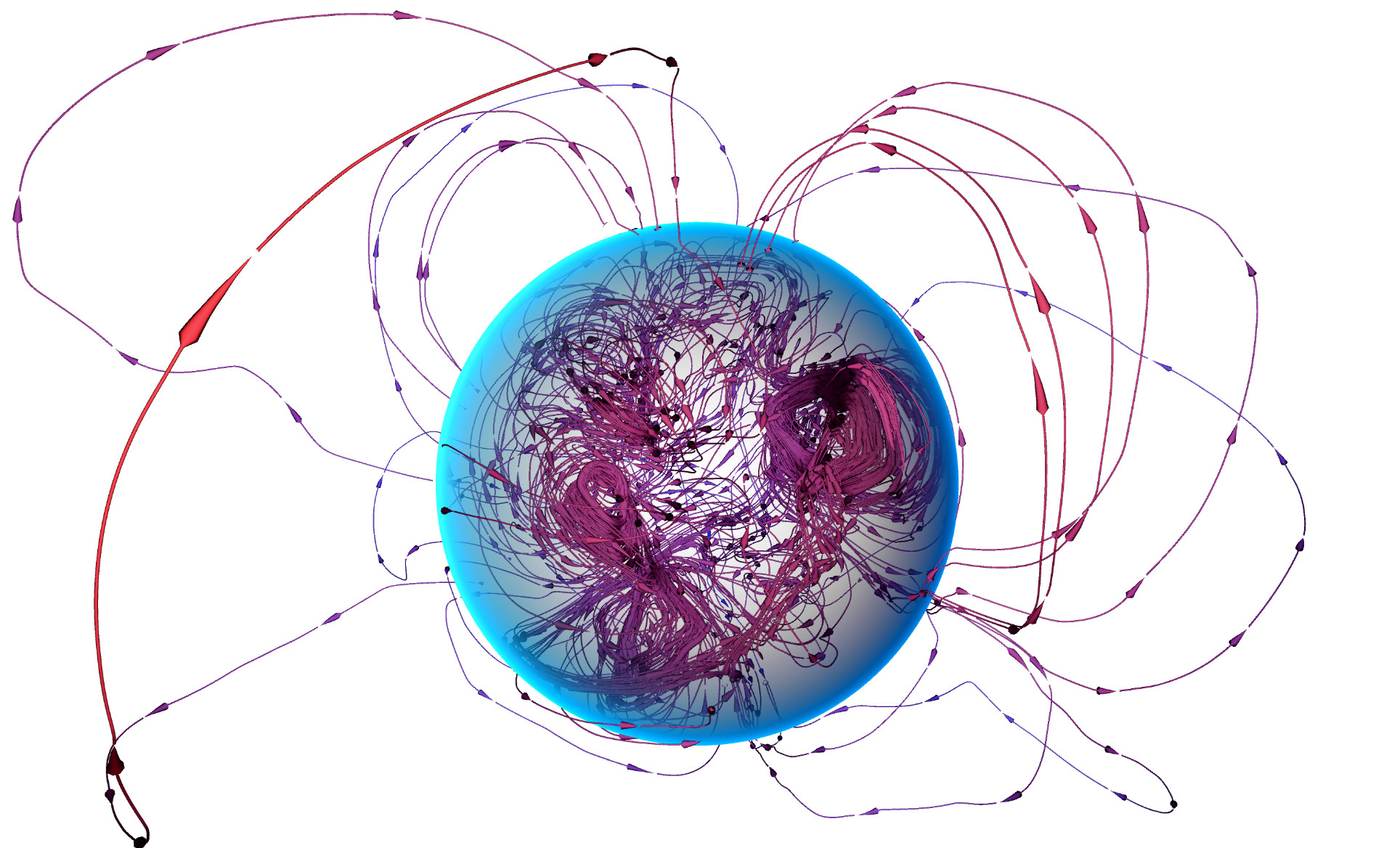}
   & \includegraphics[width=0.47\textwidth]{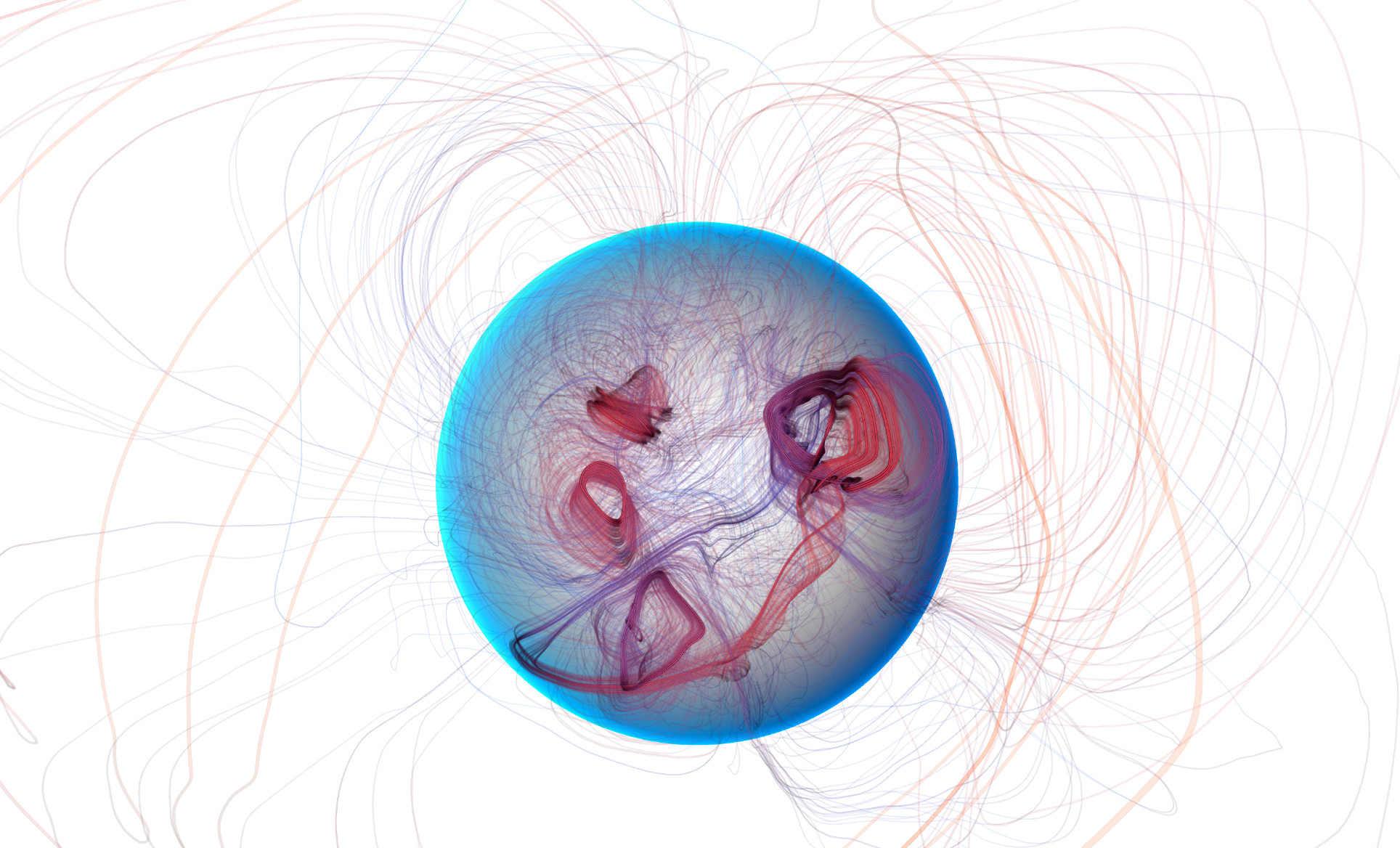}\\
  (c) Arrow streamlines & (d) Transparent streamlines\\
\end{tabular}
\caption{Four modes for the streamline visualization. (a) Streamlines as continuous lines. (b) Streamlines as dashed lines. Dashed lines are particularly interesting in motion and for recognizing vortices. (c) Streamlines as arrows. (d) Transparent streamlines. More streamlines are added in this mode to increase streamline visibility.}
\label{Fig:Modi}
\end{figure*}
\subsection{Streamline Generation}
Streamlines are curves whose tangents at all points correspond to the velocity vectors of the underlying vector field.
We use streamlines to enhance the 3D perception of the magnetic field.
Streamlines can be set specifically by the domain expert in order to investigate phenomena in isolation. 
The sparse nature of the streamlines makes it possible to work clearly and avoid occlusion effects through targeted placement. 
However, occlusion effects occur if a large number of streamlines are placed.
We have developed a fast and customizable solution to avoid occlusion effects.
Each streamline is dependent on the dataset time it is calculated.
Therefore, whenever the data changes in time, all streamlines have to be recalculated.
This happens often, as the use case of the tool is to explore different time steps of the simulation.
To satisfy the high demand of recalculations and to provide interactive streamline generation in general, we use the Unity Job System to gain multithreaded computation. 
Our streamline visualization module is structured as a pipeline in which the individual sections are calculated on different computing units.

\subsubsection{Streamline Pipeline}
The pipeline is divided into three parts: loading the dataset, constructing the curves, and animating the curves. 
In the first part, the vector data is loaded from the disk to RAM. 
This step has to be done once per time step. 
The second part calculates the streamlines on the CPU in parallel. 
A downstream compute shader then calculates curves from the raw streamlines. 
Curve properties such as tangent, normal, and curvature are calculated directly. The second part is always calculated when the generation settings change. 
This is usually the case when a new time step is selected. 
In the third part, animation properties of the curve are calculated. 
Vertices are cut out to obtain a curve pattern in a compute shader. 
The radius is also adjusted here to obtain an arrow pattern. 
In a geometry shader, the given line is then reformed into a 3D tube.
A subsequent fragment shader then shades the tubes depending on the camera angle.
Also, curve properties are color coded in the fragment shader.
The last step is carried out in every frame while the animation is running.
An overview of the pipeline is given in Fig. \ref{fig:streamlinePipeline}.

\subsubsection{Streamline Computation}

We calculate static streamlines based on a discrete vector field in 3D space.
Runge-Kutta methods are well suited to numerically approximate curves by a vector field.
Often, RK4 is used, because it yields results with a low error and low computation time compared to other line integration methods \cite{HOSEA199445}.
However, the magnitude of vectors within the visualized dataset varies up to a factor of 1000.
Therefore, the classical RK4 method with fixed step size is rather impractical, as not all magnitude ranges can be mapped equally.
Unlike LIC, each streamline has several thousand integration steps. It is therefore important to avoid error propagation and to use a more accurate integrator.
Instead RK4, we use the \textit{Runge-Kutta-Fehlberg} (RKF45) method, which is a member of the Runge-Kutta method family.
RKF45 extends the RK4 method by a computation of the 5th order Runge-Kutta and using it as an error estimation to allow a dynamic step size.
Depending on the error size, the step size is enlarged or reduced to reach a previously defined maximum error.
As suggested by Chapra and Canale \cite{chapra2010numerical}, we use the coefficients developed by Cash and Karp \cite{cash1990variable}.
The step size of RKF45 is dynamic, meaning that it changes its value during the integration.
The change of the value depends on the error, which is locally estimated using the 4th and 5th-order Runge-Kutta vectors.
As the error estimator, we use a mean absolute error (MAE) to not overcompensate for large errors.
After each iteration step, the step size $h$ is adapted to the local error:
\begin{equation}
\label{eq:stepsizeControl}
    h_{new} = \begin{cases}
        h\cdot 0.9\cdot (\frac{E_{t}}{E_{e}})^{0.2} & E_{e} \geq E_{t} \\
        h \cdot (\frac{E_{t}}{E_{e}})^{0.2} & \, \text{else}
    \end{cases}
\end{equation}
with $E_{e}$ being the computed MAE value of the current iteration (error estimation) and $E_{t}$ the maximal permitted error (error target).
If $E_e \geq E_t$, Eq. \eqref{eq:stepsizeControl} is applied and the iteration step is recalculated until $E_e$ becomes smaller than $E_t$.
Typically, the integration error is particularly large where there is a lot of turbulence. 
In such cases, the difference between the 4th and 5th order is very large. 
As the difference is large, the step size is particularly small here in order to obtain the selected error. 
This method allows finer sampling in fine turbulence structures automatically. 
If the field is locally very uniform, the difference between the 4th and 5th order is very small, so that sampling is carried out in large steps.
This saves a considerable amount of computing time, as no small step size is required to calculate the integration correctly.
For a visual representation of the RKF45-method, see Fig. \ref{Fig:RK}.

For each streamline, we integrate in both directions from the seed point on.
We keep two buffers during the generation, one buffer for each direction.
If the integration terminates, we reverse the first buffer and attach the second buffer to the end of the first buffer.
This gives us an entire buffer containing the streamline, in which the order of the points corresponds to the direction of the streamline.
We also check in every iteration step, if the computed streamline is approximately a loop.
To detect a loop, we check both ends of the current iteration.
A flag is set if both ends are moving towards each other. 
We define a merging movement as a positive scalar product of both estimated direction vectors.
If the flag is set and both ends suddenly move apart from each other, we check if the distance between both ends fall below a threshold.
If so, the generation is terminated.
The generation is also terminated if the distance between the two steps is very small.
This prevents an unnecessarily large number of points from being generated in an area where the strength of the vector field is almost zero.

We use the Unity Job System to generate the streamlines on the CPU in parallel.
The resulting buffer contains the coordinates of the streamline and is then passed on to perform further calculations on the graphics card.

\subsubsection{Streamline Post-Processing}
\begin{figure*}[t]

    \begin{subfigure}{.33\textwidth}
  \centering
    \includegraphics[width=1\linewidth]{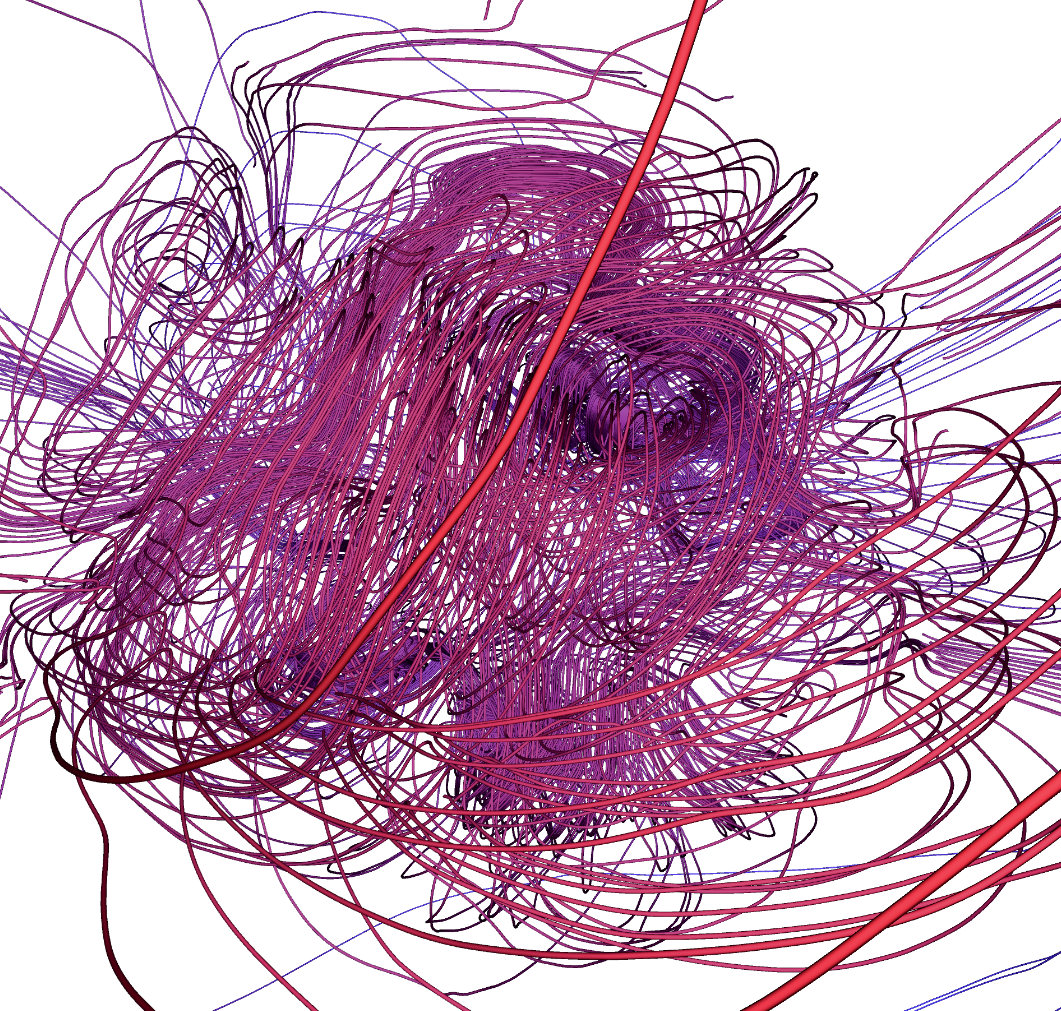}
    \caption{Full streamlines}
  \end{subfigure}%
  \begin{subfigure}{.33\textwidth}
  \centering
    \includegraphics[width=1\linewidth]{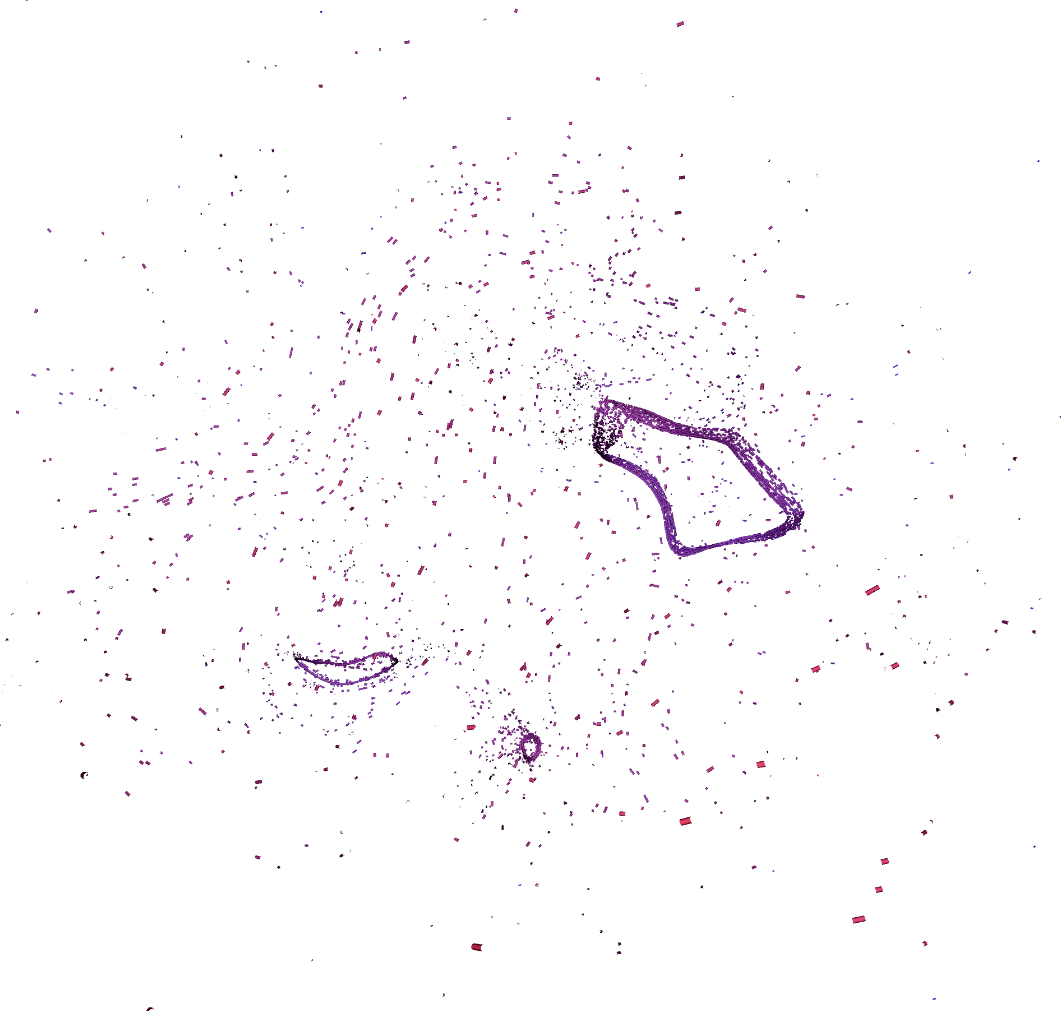}
    \caption{Dashed streamlines}
  \end{subfigure}
  \begin{subfigure}{.33\textwidth}
  \centering
    \includegraphics[width=1\linewidth]{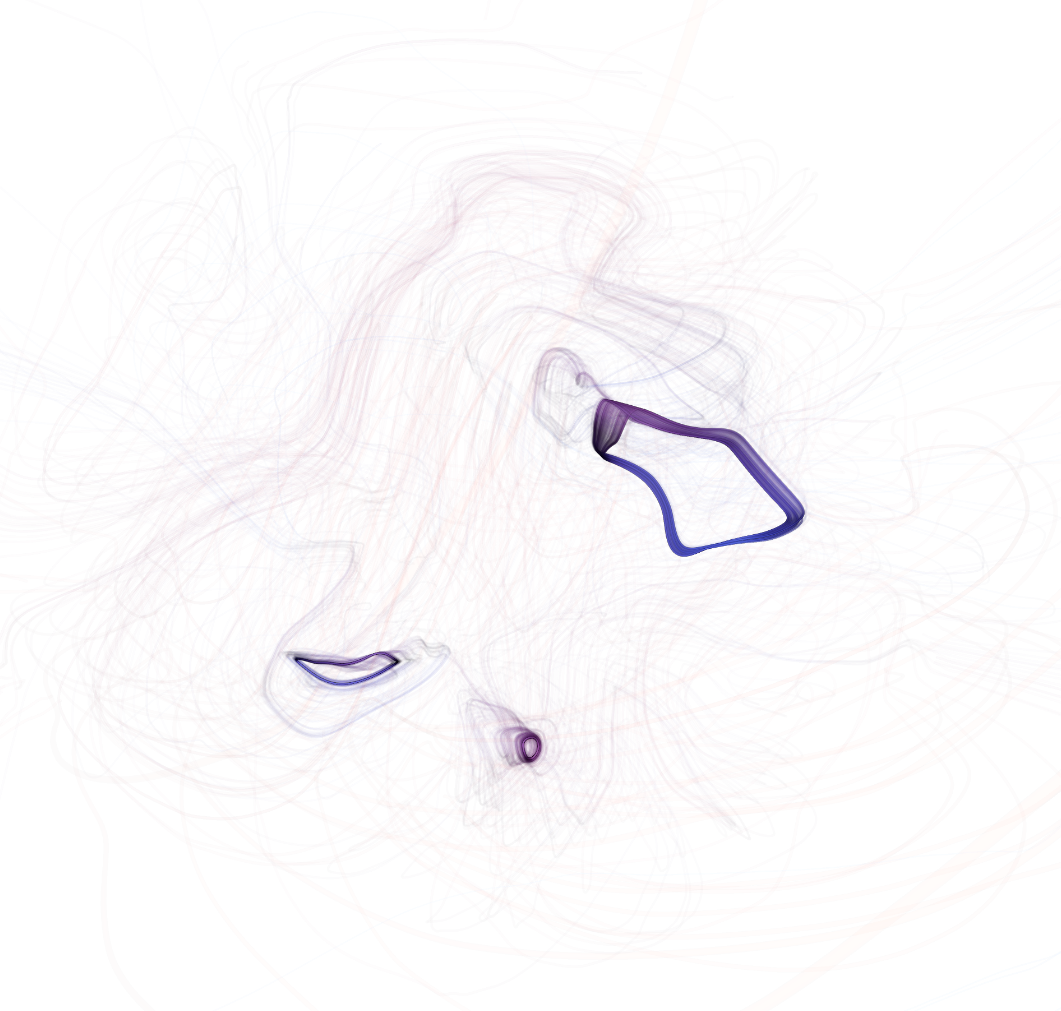}
    \caption{Transparent streamlines}
  \end{subfigure}
 \caption{Vertex isolation using dashed streamlines and transparent streamlines at time step 280. (a) Full streamlines are shown for reference. (b) Streamlines are displayed dashed with large gaps to cut the most streamlines. Streamline parts accumulate at vortices. (c) Streamlines are displayed transparently. The structure is still visible, while the vortices are clearly isolated by the accumulation of transparency.}
 \label{Fig:VortexIsolation}
\end{figure*}

After the streamline is generated on the CPU, it is passed to the graphics card inside a compute shader to transform the vertex line into a tube mesh.
Here, we calculate the normals, tangents, and bitangents for each streamline point to obtain the movement tangent frame of the streamline, using Frenet-Serret formulas \cite{mate2017frenet}.
In the geometry-shader step later, we use the tangent frame to orient the generated tubes.
In a second compute shader, which is called every frame, we compute the dashed and arrow-like appearance of the streamline.
We give users the option of displaying the streamlines as arrows to recognize the direction of the magnetic field.
Hereby, the dashed line is defined by the length of a dash segment and its subsequent gap size.
We define the sum of the gap size and the dash segment as the line segment size.
The size is defined by the number of vertices.
We interpret each vertex ID within a residual class ring modulo the line segment size.
The initial index list is constructed to follow the layout of a continuous line, where each two vertex indices define a connecting piece of the line.
For the dashed line processing, we keep the total number of vertices to leave the vertex buffer unchanged in the graphics memory.
In order to "remove" vertices anyway, i.e. to cut gaps in the line, we change the indexing of those vertices that lie in the area of the gap so that two indices point to the same vertex.
This preserves the structure of the line, as two indices each describe a line segment. 
However, the line segments within the defined gaps become invisible as their length is zero.
Now we do not have to change the number of vertices, but simply re-link the indices in each frame to create an animation.
This avoids reallocating the vertex buffer and explicitly caching vertices, saving processing time.
The animation is done by shifting each vertex id by an integer.
Outside the compute shader, the speed of the ID shifting is controlled by the user.
In areas where the overview is difficult to obtain because the vector field is very turbulent, dashed lines can be used to locate vortices. 
The line spacing is set very high so that only the streamline is barely visible. 
Streamlines accumulate at vortices, so that an increased number of line segments can be seen here. 
It is also possible to change the transparency of the streamlines so that particularly collected streamlines, and therefore vortices, are highlighted.
Transparency calculation is handled by Unity.
Although the results are more attractive to look at with transparent streamlines, they also consume more resources.
This causes vortices to separate and become visible in isolation, a comparison of both methods can be seen in Fig. \ref{Fig:VortexIsolation}.

To display the lines as cylinders and arrows, we add the radius to the properties of a streamline vertex and save it in the fourth tangent coordinate.
For the representation as a tube, the radius remains constant. 
If the streamline is to be displayed later as an arrow, the radius is adjusted according to the vertex ID within the line segment.
The radius for the arrow body remains constant and then describes a stepped tip.

For the tube generation, we use the distorted cylinders approach of Lieb \etal \cite{10.2312:vmv.20221203}. However as we do not utilize the UV mapping, we re-implement the approach inside a geometry shader so that all cylinders are calculated in parallel.
The line data constructed within the dash line shader is passed to a geometry shader.
This geometry shader gets two vertices as input and constructs a streamline aligned 3D tube as the output, using the previously calculated normal, tangent, bitangent, and radius.
Vertex pairs that lie within the gap of the dashed line collapse to length zero, as vertices within a gap are paired with itself.

Finally, the mesh data is passed to a fragment shader in which various properties are encoded by color and the tube is given a basic shading to improve visual perception.
The first color mode colors the streamline red if it is inside the neutron star and blue if it is outside. 
This makes it easy to see whether the corresponding magnetic field line is inside or outside the neutron star without having to display the neutron star directly.
In the second color mode, Pseudo Chromadepth, introduced by Ropinski \etal \cite{ropinski2006visually}, is used to improve the spatial perception of the streamlines. Pseudo Chromadepth is a widely adopted method to enhance depth visibility in various visually demanding applications \cite{hombeck2022evaluating,Meuschke_2018_VCBM,hombeck2022distance}.
The color range can be adjusted by the user to make the effect stronger or weaker. 
A comparison of the streamline visualizations is given in Fig. \ref{Fig:Modi}.

\subsubsection{Streamline animation}
Animation of streamlines helps to perceive the overall direction of the magnetic field.
We animate the streamlines by transforming the previously computed streamline into a dashed line or arrows.
Hereby, the dashed line is defined by the length of a dashed segment and its subsequent gap size.
We define the sum of the gap size and the dashed segment as the line segment size.
The size is defined by the number of vertices.
In a compute shader, we interpret each vertex id within a residual class ring modulo the line segment size.
The index list is constructed to follow the layout of a continuous line.
Therefore, each part of the mesh consists of two vertex indices.
Each vertex within the gap part of the segment is paired to itself so that it does not contribute to the overall line mesh.
The animation is done by shifting each vertex ID by an integer, which increases with a user-defined speed outside the shader.
Another time-resolved animation is created by continuously changing the time steps. During this animation, the existing seed points are not changed to ensure continuity for the user, who can freely select the time steps.

\subsection{Neutron Star Representation}
The data of the neutron star is given as a volumetric scalar field.
Each scalar indicates the density ($\rho$) of the neutron star. 
Depending on the data set, the density is time-dependent, and different density ranges can be interesting. 
Therefore, we have implemented a direct volume rendering (DVR) tool to render the neutron star density. 
However, the rendering of the neutron star by DVR is very resource-intensive. 
Since the density does not change in the tested simulation data, but the focus is set on the magnetic field, the radius of the neutron star is estimated by the density values. 
The star itself is then represented by a spherical mesh with a corresponding radius. 
We use a Fresnel shader to suggest the shape and position of the neutron star and a blue color gradient that illustrates the rotation in space, see Fig. \ref{fig:teaser}.

\subsection{Interaction \& Workflow}

The tool is created for domain expert users who explore the results of their work visually.
Two fundamental principles were taken into account during development: First, an intuitive interaction with self-explanatory controls. 
An interactive workflow is designed as follows: First, select the time step that shall be worked on. 
Second, the cross-section tool is activated to use the plane to identify points of interest. 
Alternatively, the animation of the time steps can already be switched on in order to directly find interesting points in the various time steps. 
Once an interesting area has been found by moving the cross section, the seed point tool can now be switched on. 
This allows points to be set on the cross section where streamlines are generated directly. 
Multiple points can also be set at once. 
Users can select the amount of seed points that are placed at once. Points are symmetrically placed around the center normal axis of the cross-section. We suggest placing at least 10 seed points at once and a total of 200 to utilize the best performance efficiency, see Sec. \ref{sec:technical Analysis}.

The path of the streamlines can now be examined. 
The cross section can be used again to explore new areas. 
Streamlines can be added iteratively in order to obtain an overall picture. 
By selecting the time steps, the development of the magnetic field in the corresponding areas can now also be observed.
The time steps can also be increased automatically to create an animation. During this animation, users can work freely on the tool, add seed points, move the cross-section and rotate the camera.
Some streamlines have now been set and their appearance can now be adjusted using the streamline tool.
Here, users can set the line and gap length for dashed and 
arrowed lines. The length is given in vertex count, as it directly addresses the visual appearance of the streamline mesh. Users can move a slide control to set this value to their satisfaction. Streamlines are updated on the fly in real-time.

\begin{figure*} [t]

 \includegraphics[width=1\textwidth]{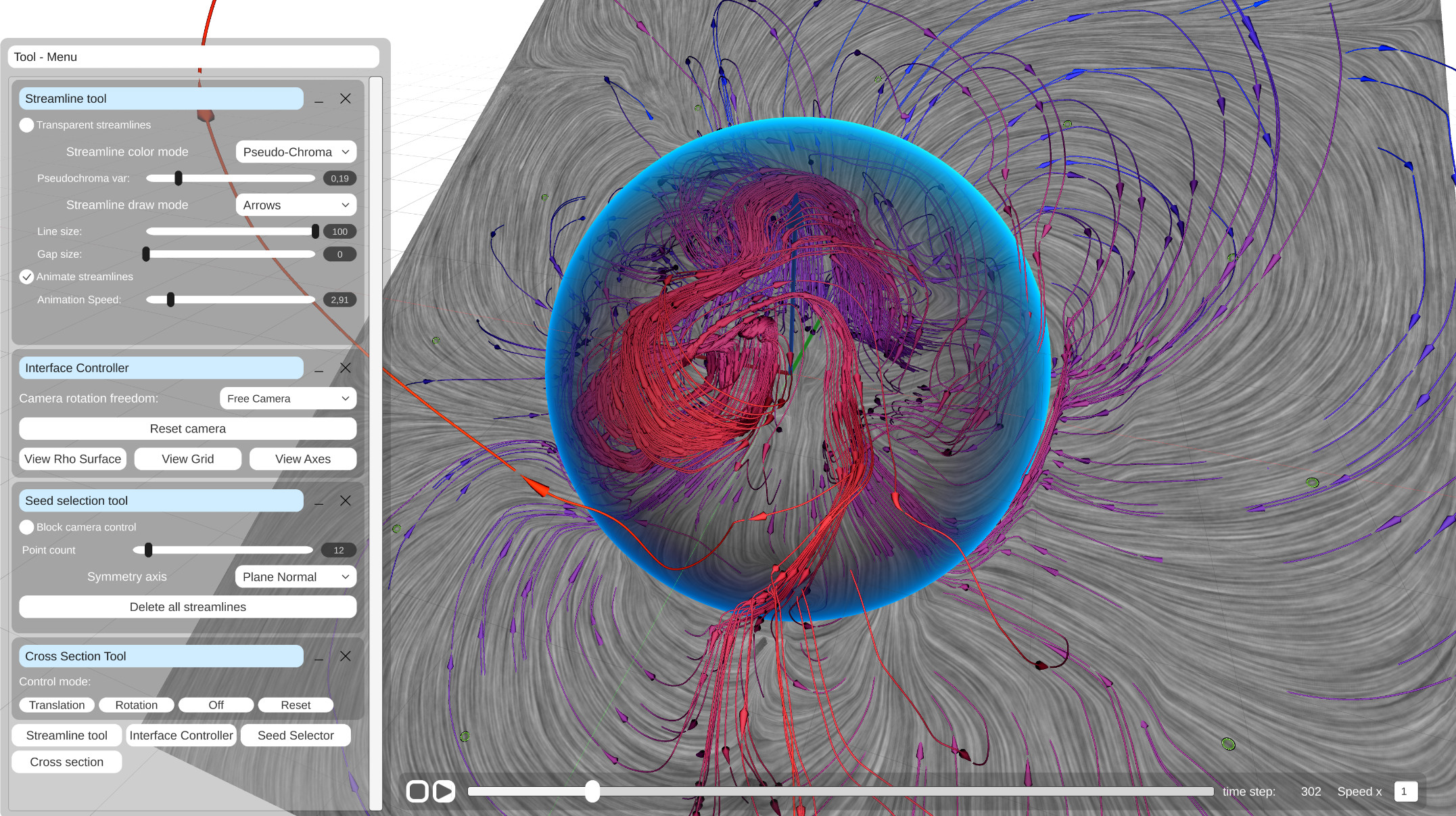}
 \caption{Screenshot of the tool with all sub-tools open, shown on the left-hand side. Individual tools can be hidden to maintain an overview. At the bottom is the time slider, which can be used to navigate through the time steps.}
 \label{Fig:Tool}
\end{figure*}

The appearance can be selected depending on the task type. 
We structure the sub-tools into distinct modules that can be turned on and off, to keep the visual interface clean. 
For an example screenshot, see Fig. \ref{Fig:Tool}.

\section{User Evaluation \& Technical Analysis}

The tool is intended to help astrophysicists to better understand the development of magnetic fields of neutron stars. In order to answer this question, we developed a qualitative study and conducted it with domain experts and a technical analysis to determine basic operability and performance.
For the qualitative study, we mainly focused on querying if requirements \textbf{R1}, \textbf{R2}, \textbf{R3}, and \textbf{R4}, discussed in Sec. \ref{stakeholder} were achieved. 
Also, we queried the main satisfaction and usability of the tool.
We performed all user queries on an Intel Core i9-12900K @ 4.4Ghz with an RTX 3090 graphics card, 32 GB RAM, and on Windows 11.

For the technical analysis, we tested the tool performance for different settings. All test settings were performed on three different hardware systems.

\subsection{User Evaluation Structure}

We subdivided the study into three sections.
In the first section, general information is requested, namely gender, age, and personal experience in the field of astrophysics, visualization, and 3D applications in general. 
A personal, anonymized key is also created in order to be able to assign participants in subsequent studies without recording personal data. 
We follow the model of P{\"o}ge \cite{poge2008personliche} to generate an anonymous key, which consists of personal variables that do not allow any conclusions to be drawn about the person, but form a unique key. 

In the second part, a practical study is carried out. 
Here the participants have to solve six tasks using the tool.
The general control system was presented in the first task. 
In the subsequent four tasks, the four main tools and their handling were presented and tested. 
Here, the participants were guided step by step through the individual interaction options of each tool and they had to perform the steps in order to continue.
The sixth task consists of three sub-tasks. 
In this section, vortex examination serves as an example, but the tool is not specifically limited to, nor exclusively designed for, finding vortices.
The participants were asked to demonstrate that they understood how to use the tool by searching for a specific vortex in the data set that disappears approximately between time steps 570 and 590 at the north pole of the neutron star.
Finding this vortex requires the use of all sub-tools. 
The second sub-task was to examine a specific time step (280). 
At this simulation time stamp, the magnetic field is very turbulent and chaotic. 
By setting many streamlines and changing the appearance to dashed lines, individual vortices can be recognized, see Fig. \ref{Fig:VortexIsolation}. 
The task was to find all three vortices at this simulation time step.
In the last sub-task, the participants had time to freely explore the tool and write down whether and what conclusions they could draw from the observation of the magnetic field.

In the third part of the evaluation survey, the impression of the tool was queried.
First, we used a standardized questionnaire about the general usability of the tool (\textit{General Usability Questionnaire}). 
Here we use the System Usability Scale (SUS) by Brooke \cite{brooke1996sus}, which is widely used to evaluate general usability and has been proven to provide significant evaluations \cite{bangor2008empirical,brooke2013sus,grier2013system,klug2017overview}.
We used the SUS also to rate each of the four sub-tools cross section, streamline tool, time slider, and seed selector.
After the general usability questionnaire, a more specific questionnaire followed that asked how effectively the tasks in the practical section were completed and how well the tool could be used to investigate specific astrophysics problems (\textit{Specific Usability Questionnaire}).
We grouped the questions in the categories \textit{User interface and control}, \textit{Visual rendering performance}, \textit{Visual information}, \textit{Visual appearance}, and \textit{Interactivity}. 
Each item of the \textit{Specific Usability Questionnaire} is a Likert-typed item ranging from strongly disagree (-{}-) to strongly agree (++). This rating is then mapped to a point scale between 0 (worst) and 4 (best). For better clarity and understanding, questions with negative implications are represented as reversed.
For a detailed breakdown of the questions, see the Suppl. material.
The questionnaire evaluates if the general requirements targeted in this paper \textbf{R1}, \textbf{R2}, \textbf{R3}, and \textbf{R4}, see Sec. \ref{stakeholder}, are fulfilled.

\subsection{Participants Profile}

We conducted the study with 6 domain experts.
Four participants were not involved in the project development process, and two were part of the stakeholder group.
Two participants stated that they had experience in astrophysics for 2 to 5 years, three for 5 to 10 years, and one for more than 10 years.
With regard to experience in Scientific Visualization, one participant stated that they had 0--1 years of experience, two participants 2--5 years, and three participants 5--10 years.
One participant stated that they had 0--1 year of experience in 3D applications. Three participants had 1--2 years of experience in 3D applications, one participant 2--5 years and one participant 5--10 years.
Their ages were between 27--34 years ($\varnothing$ 30.5 years), 5 were male, and 1 female. 
Two participants stated that they were doing their doctorate, three said they were working as a postdoc and one gave their profession as 'physicist'.
Two participants had a master's degree, four had a doctoral degree.
None had visual impairments/weaknesses.
None of the participants had worked with the tool before the evaluation.

\subsection{User Evaluation Results}

In the first practical survey, participants solved the introductory tasks for all features in 12m:39s on average, before they continued to the main tasks.
The first main task was to find the dissolution of a vortex near the north pole of the neutron star.
The time-resolved range is not clearly defined by a single time step, but is determined by an interval.
Three participants successfully found the range between time steps 570 and 583.
Two participants found a different vortex dissolving within the time interval 800--850.
There is actually another vortex in this range that we did not see during the construction of the tasks.
One participant found a vortex dissolving at time step 200.
All vortices found had not been identified before the study.
In the second practical subtask, 5 of 6 participants identified the three vortices at time step 280.
One participant discovered only one of the three vortices.
In the third subtask, which gave the participants the opportunity to examine and freely explore the tool, 2 participants documented their discoveries.
The first participant discovered the forming of a vortex coming out of the neutron star at time step 157, while most vortices stay within the neutron star.
The second participant discovered that the poles of the neutron star were moving during the simulation.

To evaluate the SUS score of the \textit{General Usability Questionnaire} we calculated the score as described by Brooke \cite{brooke1996sus}. 
Each of the ten items is assigned a value between 0 and 4. 
The values of all items are added together and multiplied by 2.5.
The resulting SUS scores are categorized using the work of Bangor \etal \cite{bangor2009determining}.
For the detailed SUS evaluation, see the Suppl. material, where the results are broken down by feature, participant, and questions. 
The overall SUS score, which represents the main assessment of the tool in general, is 96.7\%, which corresponds to a rating of \textit{Excellent} to \textit{Best Imaginable}.
The sub-tools were also rated \textit{Excellent} to \textit{Best Imaginable}, however, with different numerical ratings:
The SUS score for the 'seed selector' feature reached the value 90\%, for the 'streamline tool' feature 95.4\%, for the 'time slider' feature 95.4\%, for the 'cross section' feature 95.8\%.

The \textit{Specific Usability Questionnaire} was rated with 88.4\% (467 out of 528 points).
However, this value is not on the SU scale and should be used as a quality estimate, especially to determine differences in the individual evaluation categories.
The rating is based on the ratings of the categories as follows:
\textit{User interface and control}: 81.9\% (59 of 72), \textit{Visual rendering performance}: 86.1\% (62 of 72), \textit{Visual information}: 88.9\% (128 of 144), \textit{Visual appearance}: 92.7\% (89 of 96), and \textit{Interactivity}: 89.6\% (129 of 144).
For the detailed \textit{Specific Usability Questionnaire} evaluation results, see Sec.~\ref{Sec:App}.

All participants commented and shared their impressions along with suggestions in the free response section.
We have compiled the points and summarized them as follows:

\begin{itemize}
    \item \textit{"It would be useful to switch between data sets.}
    \item \textit{The time slider should be changeable in speed and certain time steps should be selectable individually.}
    \item \textit{More control of the position and orientation of the cross section and streamline placement (to be able to control the position and orientation via code).}
    \item \textit{It would be useful to group streamlines.}
    \item \textit{The ability to seed streamlines from any plane in real-time was very useful.}
    \item \textit{The visualization of the magnetic field is very intuitive and easy to use."}
\end{itemize}

\subsection{Technical Analysis}
\label{sec:technical Analysis}
For the technical test runs, we used three different hardware systems ([S1], [S2], [S3]) with ascending computation power (for details see the Suppl. material).
We conducted several performance tests on [S1], [S2], and [S3] testing varying amounts of streamlines across different time steps.
All tests were repeated 9 times to ensure reliability.
Our analysis revealed that the streamline calculation was most efficient per integration when performing 370,000 integration steps across all systems: 46ms [S1], 23ms [S2], and 13ms [S3] total calculation time.
This amount of integration steps corresponds to 100 to 200 streamlines, depending on the streamline length.
The loading of the first LOD data takes 41ms [S1], 29ms [S2], and 26ms [S3].
Note that the data loading and streamline calculation run in parallel.
The average FPS count for animated streamlines did not drop below 60 in our tests, which end at 44,000,000 triangles.
The minimum number of frames (27 FPS) is reached at 44,000,000 triangles in [S2].
For detailed numbers, see the Suppl. material.

\section{Discussion}

The practical part of the evaluation gave the participants the opportunity to familiarize themselves with the tool. 
It took the participants an average of 12 minutes to familiarize themselves with the tool and then solve complex tasks. 
Three experts found the correct time interval in which the vortex disappeared at the north pole of the star, two found a similar vortex.
Five experts were able to find all three vortices at time 280.
Although none of the participants had worked with the tool before, two participants were able to gain knowledge during the evaluation: One discovered a vortex located outside the neutron star and the other found that the poles shifted during the simulation.
The quick familiarization and insightful use of the tool show that it is very well suited to the work of the domain experts.
This impression is reinforced by the excellent result of the SUS questionnaire (\textit{General Usability Questionnaire}) of 96.7\%. 
As the individual parts of the tool were also rated as excellent, we see the usefulness of the individual components in combination as confirmed.

\textbf{R1: Interactive}.
The interactivity was rated on average with 89.9\%.
Here, one participant had problems with the accuracy of setting Seed Points.
The participant suggested setting Seed Points by code, for example, by entering the exact coordinates.
This is certainly a good idea in order to set the seed points even more precisely.
Other participants appreciated the fact that seeding points are easy to set anywhere in the 3D area.
Despite the potential for improvement, based on the remaining ratings the tool provides a high degree of interactivity.

\textbf{R2: Reactive}.
With an average rating of 86.2\%, the visual rendering performance has been rated very high. 
As expected, some participants experienced stuttering, which occurs during frame changes depending on the number of streamlines.
However, the high visual rendering performance rating and more than 60 fps in average use show that the tool is reactive.

\textbf{R3: Intuitive}.
The tool was explicitly complimented on the fact that, unlike similar tools, it is very intuitive in use. 
This is confirmed on the one hand by the short time required for the tasks and on the other hand by the good evaluation of the controls.
The user interface and control were rated on average with 81.9\% and claimed to be easy to use.

\textbf{R4: Engaging}.
In addition to the interaction options, the transfer of information and its presentation were also rated positively. 
The high ratings of the visual information 88.9\% and visual appearance 92.7\% show that the tool can be used both for gaining knowledge and for conveying information for outreach.

Due to the excellent overall usability ratings and the good ratings of the individual criteria, we consider the requirements (\textbf{R1, R2, R3, R4}, see Sec. \ref{stakeholder}) to be fulfilled.
Nevertheless, there is also criticism to be discussed: The goal was to create an intuitive control of all exploration elements.
The intuitiveness of the streamline placement is given by the simple use of the cross section and the placement of the seed points on this plane.
However, it should also be possible to control the elements through precise code input. 
It was noted that it would be good if the speed of the time slider could be adjusted and individual time steps could be selected more precisely.
A control of the time slider, which makes it possible to regulate the speed and select specific time steps, has already been implemented after the evaluation.
For the data used, it is true that all datasets with the structure described in the paper can be used, but the tool lacks an integrated pre-processing function. 
This takes place in an external Python script.
The grouping of streamlines would be a promising extension that can be discussed and implemented in the future.
\section{Conclusion \& Future Work}

We present a novel visualization tool, created to help astrophysicists to explore the evolution of magnetic fields in neutron star simulations.
The development was carried out in regular consultation with domain experts in order to meet their needs. 
For this purpose, four interaction techniques were developed, the interplay of which enables interactive, reactive, intuitive, and attractive exploration of the magnetic field data.
We combine a dense vector field visualization using an interactively movable cross section with sparse vector field visualization using specifically placed streamlines.
The cross section gives users the option of placing streamlines at selected locations in order to examine specific areas.
In addition, the time-variant data set can be explored in real-time using the intuitive time slider.
The implementation of the features leverages modern graphics hardware to enable real-time interaction with massive data sets.
We show that the tool provides valuable insights into the astrophysical study of the evolution of magnetic fields of neutron stars.
The visual results are suitable for informative outreach.
The evaluation of the tool has shown that it can be used very well for the everyday research needs of astrophysicists. 
In the future, we intend to integrate streamline clustering so that different areas may be compared.
It should also be possible to further specify the position of the streamlines by manually entering the seed point coordinates or to set a series of seed points whose coordinates are described by a function.

\newpage
\bibliographystyle{unsrt}
\bibliography{references}  

\newpage
\section{Appendix}
\label{Sec:App}
\definecolor{++}{rgb}{0.6,0.6,1}
\definecolor{+}{rgb}{0.8,0.8,1}
\definecolor{o}{rgb}{1,1,1}
\definecolor{-}{rgb}{1,0.8,0.8}
\definecolor{--}{rgb}{1,0.6,0.6}
\newcommand{\Evpp}{\cellcolor{++}\texttt{++}}
\newcommand{\Evp}{\cellcolor{+}\texttt{+}}
\newcommand{\Evo}{\cellcolor{o}\texttt{o}}
\newcommand{\Evm}{\cellcolor{-}\texttt{-}}
\newcommand{\Evmm}{\cellcolor{--}\texttt{-{}-}}
\newcommand{\EvppR}{\cellcolor{++}\texttt{-{}-}}
\newcommand{\EvpR}{\cellcolor{+}\texttt{-}}
\newcommand{\EvoR}{\cellcolor{o}\texttt{o}}
\newcommand{\EvmR}{\cellcolor{-}\texttt{+}}
\newcommand{\EvmmR}{\cellcolor{--}\texttt{++}}
\newcommand{\explen}{3.05cm}
\newcommand{\EvolOverall}{Overall}
\newcommand{\EvolSeed}{Seed Selection}
\newcommand{\EvolVector}{Streamline}
\newcommand{\EvolTime}{Time-Slider}
\newcommand{\EvolCrossSection}{Cross-Section}

\begin{table}[h]

    \caption{
Overview of the answers of the specific part of the evaluation. 
Results are broken down by feature and participant (\textit{Par.}).
The answer options vary from strongly disagree (-{}-) to strongly agree (++). 
Colors indicate how positive the answer is, ranging from red (very negative) to blue (very positive).}
\renewcommand{\arraystretch}{0.95}
    \begin{tabularx}{\linewidth}{|X|c|c|c|c|c|c|}
        \multicolumn{1}{c|}{} & \rotatebox{90}{Par. 1}& \rotatebox{90}{Par. 2}& \rotatebox{90}{Par. 3}& \rotatebox{90}{Par. 4}& \rotatebox{90}{Par. 5}& \rotatebox{90}{Par. 6}\\
    \hline
    \multicolumn{7}{|c|}{User interface and control} \\
    \hline
    The user interface was clearly arranged. & \Evp & \Evpp & \Evpp & \Evp & \Evp & \Evo \\
    \hline
    Controlling the camera was intuitive. & \Evp & \Evpp & \Evpp & \Evo & \Evpp & \Evp \\
    \hline
    The interface was pleasant to use. & \Evp & \Evpp & \Evpp & \Evp & \Evpp &  \Evo\\
    \hline
    \multicolumn{7}{|c|}{Visual rendering performance} \\
    \hline
    The rendering felt smooth and responsive. & \Evpp & \Evp &  \Evp& \Evpp & \Evpp &  \Evo\\
    \hline
    I have experienced significant lags (stuttering, long loading times). & \EvppR & \EvoR & \EvpR & \EvppR & \EvppR & \EvoR \\
    \hline
    I often had to wait for the program to finish before I could continue. & \EvppR & \EvppR &\EvppR & \EvppR & \EvppR & \EvpR \\
    \hline
    \multicolumn{7}{|c|}{Visual information} \\
    \hline
    I was able to gain information from the visual representation of the data. & \Evpp & \Evpp &\Evpp & \Evp & \Evpp &  \Evo\\
    \hline
    The cross section helped me to understand the structure of the magnetic field. & \Evpp & \Evpp & \Evpp& \Evpp & \Evpp & \Evp \\
    \hline
    I was able to look at the data set in a new way. & \Evpp & \Evp & \Evp & \Evm & \Evpp & \Evp \\
    \hline
    The streamlines helped me to understand the structure of the magnetic field. & \Evpp & \Evpp & \Evpp & \Evp & \Evpp & \Evp \\
    \hline
    The arrow setting for streamlines helped me understanding the direction of the vector field. & \Evp & \Evpp & \Evpp & \Evpp& \Evpp & \Evp \\
    \hline
    The animation of the time steps helped me to better understand the development of the magnetic field. & \Evpp & \Evpp& \Evp &\Evpp & \Evpp &  \Evp\\
    \hline
    \multicolumn{7}{|c|}{Visual appearance} \\
    \hline
    I believe the visualization helps to convey knowledge. & \Evpp & \Evpp& \Evpp & \Evpp & \Evpp & \Evp \\
    \hline
    The visualization looks pleasant. & \Evpp & \Evpp& \Evpp & \Evpp & \Evpp & \Evp \\
    \hline
    The animated streamlines helped to understand the overall data structure. & \Evpp & \Evpp& \Evpp & \Evo & \Evpp & \Evp \\
    \hline
    The streamline visualization helped me better understand the data. & \Evpp & \Evpp& \Evpp & \Evp & \Evpp & \Evp \\
    \hline
    \multicolumn{7}{|c|}{Interactivity} \\
    \hline
    There were too much settings to adjust. & \EvppR & \EvppR & \EvppR & \EvppR & \EvppR & \EvppR \\
    
    \hline
    The cross section helped me to set seed points correctly. & \Evpp & \Evp &\Evpp & \Evm & \Evpp &  \Evp\\
    \hline
    I found the ability to look at different time steps useful. & \Evpp & \Evpp & \Evp & \Evpp& \Evpp & \Evpp \\
    \hline
    I found the streamline appearance options useful. & \Evo & \Evpp & \Evpp & \Evpp & \Evpp & \Evp \\
    \hline
    I could navigate well with the cross section. & \Evpp & \Evpp& \Evp & \Evpp & \Evpp & \Evp \\
    \hline
    The manual point selection helped me to investigate my task in more detail. & \Evpp & \Evpp & \Evpp & \Evm & \Evpp & \Evp \\
    \hline
    \end{tabularx}
    \label{tab:Exteval}
\end{table}

\begin{table}
\caption{Overview of the answers of the SUS part of the evaluation. 
Results are broken down by feature and participant (\textit{Par.}).
The answer options vary from strongly disagree (-{}-) to strongly agree (++). 
Colors indicate how positive the answer is, ranging from red (very negative) to blue (very positive).}
\renewcommand{\arraystretch}{0.985}
\begin{tabularx}{\linewidth}{|X|l|c|c|c|c|c|c|}
    \multicolumn{2}{c|}{} & \rotatebox{90}{Par. 1}& \rotatebox{90}{Par. 2}& \rotatebox{90}{Par. 3}& \rotatebox{90}{Par. 4}& \rotatebox{90}{Par. 5}& \rotatebox{90}{Par. 6}\\
    \hline
    \multirow{5}{\explen}{I think that I would like to use this system frequently.} 
     &\EvolOverall &\Evpp  &\Evpp &\Evpp  &\Evpp  &\Evpp  &\Evp \\
     \cline{2-8}
     &\EvolSeed &\Evpp  &\Evp &\Evpp  &\Evp  &\Evpp  &\Evp \\
     \cline{2-8}
     &\EvolVector &\Evpp  &\Evp &\Evpp  &\Evpp  &\Evpp  &\Evo \\
     \cline{2-8}
     &\EvolTime &\Evpp  &\Evpp &\Evpp  &\Evp  &\Evpp  &\Evpp \\
     \cline{2-8}
     &\EvolCrossSection &\Evpp  &\Evpp &\Evp  &\Evpp  &\Evpp  &\Evp \\
    \hline
    \multirow{5}{\explen}{I found the system unnecessarily complex.} 
     &\EvolOverall &\EvppR  &\EvppR &\EvppR  &\EvppR  &\EvppR  &\EvpR \\
     \cline{2-8}
     &\EvolSeed &\EvppR  &\EvppR &\EvppR  &\EvppR  &\EvppR  &\EvpR \\
     \cline{2-8}
     &\EvolVector &\EvppR  &\EvppR &\EvppR  &\EvppR  &\EvppR  &\EvpR \\
     \cline{2-8}
     &\EvolTime &\EvppR  &\EvppR &\EvppR  &\EvppR  &\EvppR  &\EvppR \\
     \cline{2-8}
     &\EvolCrossSection &\EvppR  &\EvppR &\EvppR  &\EvppR  &\EvppR  &\EvpR \\
    \hline
    \multirow{5}{\explen}{I thought the system was easy to use.} 
     &\EvolOverall &\Evpp  &\Evpp &\Evpp  &\Evpp  &\Evpp  &\Evp \\
     \cline{2-8}
     &\EvolSeed &\Evpp  &\Evpp &\Evpp  &\Evo  &\Evpp  &\Evp \\
     \cline{2-8}
     &\EvolVector &\Evpp  &\Evpp &\Evpp  &\Evpp  &\Evpp  &\Evp \\
     \cline{2-8}
     &\EvolTime &\Evpp  &\Evpp &\Evpp  &\Evp  &\Evpp  &\Evp \\
     \cline{2-8}
     &\EvolCrossSection &\Evpp  &\Evpp &\Evpp  &\Evpp  &\Evpp  &\Evp \\
    \hline
    \multirow{5}{\explen}{I think that I would need the support of a technical person to be able to use this system.} 
     &\EvolOverall &\EvppR  &\EvppR &\EvppR  &\EvppR  &\EvppR  &\EvppR \\
     \cline{2-8}
     &\EvolSeed &\EvppR  &\EvppR &\EvppR  &\EvppR  &\EvppR  &\EvppR \\
     \cline{2-8}
     &\EvolVector &\EvppR  &\EvppR &\EvppR  &\EvppR  &\EvppR  &\EvppR \\
     \cline{2-8}
     &\EvolTime &\EvppR  &\EvppR &\EvppR  &\EvppR  &\EvppR  &\EvppR \\
     \cline{2-8}
     &\EvolCrossSection &\EvppR  &\EvppR &\EvppR  &\EvppR  &\EvppR  &\EvppR \\
    \hline
    \multirow{5}{\explen}{I found the various functions in this system were well integrated.} 
     &\EvolOverall &\Evpp  &\Evpp &\Evp  &\Evpp  &\Evp  &\Evpp \\
     \cline{2-8}
     &\EvolSeed &\Evpp  &\Evpp &\Evp  &\Evp  &\Evo  &\Evo \\
     \cline{2-8}
     &\EvolVector &\Evpp  &\Evpp &\Evp  &\Evpp  &\Evp  &\Evp \\
     \cline{2-8}
     &\EvolTime &\Evpp  &\Evpp &\Evo  &\Evpp  &\Evp  &\Evp \\
     \cline{2-8}
     &\EvolCrossSection &\Evpp  &\Evpp &\Evp  &\Evpp  &\Evpp  &\Evo \\
    \hline
    \multirow{5}{\explen}{I thought there was too much inconsistency in this system.} 
     &\EvolOverall &\EvppR  &\EvppR &\EvppR  &\EvppR  &\EvppR  &\EvpR \\
     \cline{2-8}
     &\EvolSeed &\EvppR  &\EvppR &\EvppR  &\EvpR  &\EvpR  &\EvppR \\
     \cline{2-8}
     &\EvolVector &\EvppR  &\EvppR &\EvppR  &\EvppR  &\EvppR  &\EvppR \\
     \cline{2-8}
     &\EvolTime &\EvppR  &\EvppR &\EvppR  &\EvppR  &\EvpR  &\EvppR \\
     \cline{2-8}
     &\EvolCrossSection &\EvppR  &\EvppR &\EvppR  &\EvppR  &\EvppR  &\EvpR \\
    \hline
    \multirow{5}{\explen}{I would imagine that most people would learn to use this system very quickly.} 
     &\EvolOverall &\Evpp  &\Evpp &\Evpp  &\Evpp  &\Evpp  &\Evp \\
     \cline{2-8}
     &\EvolSeed &\Evpp  &\Evpp &\Evpp  &\Evp  &\Evpp  &\Evp \\
     \cline{2-8}
     &\EvolVector &\Evpp  &\Evpp &\Evpp  &\Evpp  &\Evpp  &\Evp \\
     \cline{2-8}
     &\EvolTime &\Evpp  &\Evpp &\Evpp  &\Evpp  &\Evpp  &\Evp \\
     \cline{2-8}
     &\EvolCrossSection &\Evpp  &\Evpp &\Evpp  &\Evpp  &\Evpp  &\Evp \\
    \hline
    \multirow{5}{\explen}{I found the system very cumbersome to use.} 
     &\EvolOverall &\EvppR  &\EvppR &\EvppR  &\EvppR  &\EvppR  &\EvppR \\
     \cline{2-8}
     &\EvolSeed &\EvppR  &\EvppR &\EvppR  &\EvppR  &\EvppR  &\EvpR \\
     \cline{2-8}
     &\EvolVector &\EvppR  &\EvppR &\EvppR  &\EvppR &\EvppR  &\EvpR \\
     \cline{2-8}
     &\EvolTime &\EvppR  &\EvppR &\EvppR  &\EvppR  &\EvppR  &\EvppR \\
     \cline{2-8}
     &\EvolCrossSection &\EvppR  &\EvppR &\EvppR  &\EvppR  &\EvppR  &\EvpR \\
    \hline
    \multirow{5}{\explen}{I felt very confident using the system.} 
     &\EvolOverall &\Evpp  &\Evpp &\Evpp  &\Evp  &\Evpp  &\Evpp \\
     \cline{2-8}
     &\EvolSeed &\Evpp  &\Evpp &\Evpp  &\Evm  &\Evpp  &\Evp \\
     \cline{2-8}
     &\EvolVector &\Evpp  &\Evpp &\Evpp  &\Evpp  &\Evpp  &\Evp \\
     \cline{2-8}
     &\EvolTime &\Evpp  &\Evpp &\Evpp  &\Evp  &\Evpp  &\Evpp \\
     \cline{2-8}
     &\EvolCrossSection &\Evpp  &\Evpp &\Evpp  &\Evpp  &\Evpp  &\Evpp \\
    \hline
    \multirow{5}{\explen}{I needed to learn a lot of things before I could get going with this system.} 
     &\EvolOverall &\EvppR  &\EvppR &\EvppR  &\EvppR  &\EvppR  &\EvppR \\
     \cline{2-8}
     &\EvolSeed &\EvpR   &\EvppR &\EvppR  &\EvppR  &\EvppR  &\EvpR \\
     \cline{2-8}
     &\EvolVector &\EvppR  &\EvppR &\EvppR  &\EvppR  &\EvppR  &\EvppR \\
     \cline{2-8}
     &\EvolTime &\EvpR   &\EvppR &\EvppR  &\EvppR  &\EvppR  &\EvppR \\
     \cline{2-8}
     &\EvolCrossSection &\EvppR  &\EvppR &\EvppR  &\EvppR  &\EvppR  &\EvppR \\
    \hline
\end{tabularx}
\label{tab:SUSeval}
\end{table}

\begin{table*}
\caption{
System configurations used during the performance tests.}
\begin{tabular}{|c||c|c|c|c|c|c|c|}
\hline
System & OS & GPU & GPU Memory & Processor & Cores & RAM & SSD Read\\ \hhline{|=#=|=|=|=|=|=|=|}
System 1 [S1] &Windows 10  & GTX 1080 Ti & 11 GB & i7-7700K @ 4.2 GHz & 8 & 16 GB & 560 MB/s\\ \hline
System 2 [S2] &Windows 11 & RTX 2070 SUPER & 8 GB & i9-9900KF @ 3.6 GHz& 16 & 64 GB & 560 MB/s \\ \hline
System 3 [S3] & Windows 11 & RTX 3090 & 24 GB &  i9-12900K @ 3.2 GHz & 24 & 32 GB& 560 MB/s\\ \hline
\end{tabular}
\label{tab:SystemConfig}
\end{table*}

\begin{table*}
\caption{
Results from the performance tests on different systems (S1).}
\begin{tabularx}{\linewidth}{|X||c|c|c|c|c|c|c|c|}
\hline
\multicolumn{9}{|c|}{\textbf{System 1 [S1]}} \\ \hhline{|=#=|=|=|=|=|=|=|=|}
Streamline Count & 10 & 20 & 50 & 100 & 200 & 500 & 1000 & 1500 \\ \hline
Avg. Integrations & 36 730 & 72 776 & 182 922 & 367 697 & 736 777 & 1 841 085 & 3 669 414 & 5 519 896 \\ \hline
Avg. Triangles & 293 831 & 582 206 & 1 463 377 & 2 941 574 & 5 894 215 & 14 728 681 & 29 355 309 & 44 159 165 \\ \hline
Avg. Gen. Time (ms) & 6.13 & 9.82 & 22.70 & 45.54 & 98.91 & 259.10 & 578.71 & 2137.49 \\ \hline
Avg. Gen. Time (ms)\newline per 10k Integrations & 1.77 & 1.46 & 1.35 & 1.35 & 1.41 & 1.44 & 1.71 & 4.48 \\ \hline
Avg. FPS & 1 559 & 1 231 & 860 & 551 & 329 & 166 & 90 & 61 \\ \hline
Min. FPS & 1 372 & 996 & 614 & 360 & 191 & 86 & 45 & 30 \\ \hline
Max. FPS & 1 997 & 1 825 & 1 471 & 1 131 & 731 & 398 & 220 & 150 \\ \hline
\end{tabularx}
\label{tab:PerformanceTest}
\end{table*}

\begin{table*}
\caption{
Results from the performance tests on different systems (S2).}
\begin{tabularx}{\linewidth}{|X||c|c|c|c|c|c|c|c|}
\hline \hline
\multicolumn{9}{|c|}{\textbf{System 2 [S2]}} \\ \hhline{|=#=|=|=|=|=|=|=|=|}
Streamline Count & 10 & 20 & 50 & 100 & 200 & 500 & 1000 & 1500 \\ \hline
Avg. Integrations & 36 772 & 73 407 & 183 393 & 368 028 & 735 017 & 1 842 943 & 3 684 459 & 5 517 475 \\ \hline
Avg. Triangles & 294 179 & 587 255 & 1 467 143 & 2 944 220 & 5 880 133 & 14 743 543 & 29 475 670 & 44 139 797 \\ \hline
Avg. Gen. Time (ms) & 3.70 & 6.21 & 11.88 & 23.07 & 67.82 & 152.61 & 425.43 & 996.32 \\ \hline
Avg. Gen. Time (ms)\newline per 10k Integrations & 1.18 & 0.94 & 0.73 & 0.71 & 0.95 & 0.86 & 1.31 & 2.00 \\ \hline
Avg. FPS & 1 574 & 1 238 & 850 & 564 & 343 & 179 & 90 & 64 \\ \hline
Min. FPS & 1 402 & 1017 & 611 & 350 & 188 & 79 & 40 & 27 \\ \hline
Max. FPS & 2 016 & 1 855 & 1 527 & 1 206 & 803 & 493 & 242 & 179 \\ \hline
\end{tabularx}
\label{tab:PerformanceTest}
\end{table*}

\begin{table*}
\caption{
Results from the performance tests on different systems (S3).}
\begin{tabularx}{\linewidth}{|X||c|c|c|c|c|c|c|c|}
\hline \hline
\multicolumn{9}{|c|}{\textbf{System 3 [S3]}} \\ \hhline{|=#=|=|=|=|=|=|=|=|}
Streamline Count & 10 & 20 & 50 & 100 & 200 & 500 & 1000 & 1500 \\ \hline
Avg. Integrations & 36 721 & 73 709 & 183 092 & 367 415 & 736 715 & 1 843 537 & 3 678 072 & 5 517 834 \\ \hline
Avg. Triangles & 293 775 & 589 676 & 1 464 738 & 2 939 327 & 5 893 727 & 14 748 303 & 29 424 580 & 44 142 677 \\ \hline
Avg. Gen. Time (ms) & 2.73 & 3.20 & 6.98 & 12.81 & 44.47 & 100.81 & 230.16 & 522.24 \\ \hline
Avg. Gen. Time (ms)\newline per 10k Integrations & 0.90 & 0.55 & 0.47 & 0.43 & 0.61 & 0.57 & 0.70 & 1.10 \\ \hline
Avg. FPS & 2 544 & 2 094 & 1 399 & 1 031 & 646 & 315 & 172 & 118 \\ \hline
Min. FPS & 2 278 & 1 748 & 1 003 & 667 & 369 & 158 & 82 & 56 \\ \hline
Max. FPS & 3 130 & 2 930 & 2 521 & 2 024 & 1 447 & 776 & 439 & 301 \\ \hline
\end{tabularx}
\label{tab:PerformanceTest}
\end{table*}

\begin{table*}
\centering
\caption{
Results from the data loading performance tests on different \\systems. Values are given in milliseconds (ms).}
\begin{tabular}{|p{3.1cm}||r|r|r|}
\hline
System & LOD 0 & LOD 1 & LOD 2 \\ \hhline{|=#=|=|=|}
System 1 [S1] &41,49  & 114,76 & 564,39 \\ \hline
System 2 [S2] &29.09 & 74.82 & 172,39  \\ \hline
System 3 [S3] & 25,76 & 51,44 & 141,48  \\ \hline
\end{tabular}
\label{tab:MemoryTest}
\end{table*}

\end{document}